\documentstyle[aps,psfig,multicol]{revtex}
\begin{document}
\draft
\widetext
\title{
Schwinger-Keldysh Approach to Disordered and Interacting Electron Systems:
\\
Derivation of Finkelstein's Renormalization Group Equations}

\author{Claudio Chamon$^{1,2}$, Andreas W.W. Ludwig$^{3,4}$,
and Chetan Nayak$^{4,5}$}
\address{
$^1$Department of Physics,
University of Illinois at Urbana-Champaign,
Urbana, IL 61801-3080\\
$^2$Department of Physics,
Boston University,
Boston MA 02215\\
$^3$Physics Department, University of California, Santa Barbara,
CA 93106-4030\\
$^4$Institute for Theoretical Physics, University of California,
Santa Barbara, CA 93106-4030\\
$^5$Physics Department, University of California, Los Angeles
CA 90095-1547
}
\maketitle
\begin{abstract}
We develop a dynamical approach based on the Schwinger-Keldysh
formalism to derive a field-theoretic description of disordered and
interacting electron systems. We calculate within this formalism the
perturbative RG equations for interacting electrons expanded around a
diffusive Fermi liquid fixed point, as obtained originally by
Finkelstein using replicas.   The major simplifying feature
of this approach, as compared to Finkelstein's is  
that instead of $N \to 0$ replicas, we only need to consider $N=2$ species.
We compare the dynamical Schwinger-Keldysh approach and the
replica methods, and we present a simple and pedagogical RG procedure
to obtain Finkelstein's RG equations.

\end{abstract}

\pacs{PACS: 71.10 -w, 71.23.An, 71.30 +h}

\section{INTRODUCTION}
\label{sec:intro}

Since the original idea of an impurity driven metal-insulator
transition (MIT) was put forward by Anderson\cite{Anderson}, a
substantial amount of work has been carried out to understand this
problem in the language of phase transitions and of the
Renormalization Group (RG) \cite{Thouless,Wegner,Gang4}. Simple
scaling arguments were made for the case of non-interacting electrons.
It was demonstrated that there would always be a metal-to-insulator
transition in three dimensions as a function of the disorder strength,
whereas in one and two dimensions even the weakest amount of disorder
would make the system an insulator at zero temperature
\cite{Gang4}. The lower critical dimension for the MIT is $d=2$ for
the non-interacting electron problem, and $\epsilon=d-2$ expansions
have been carried out to determine the critical exponents
characterizing the phase transition \cite{Wegner}.

These scaling ideas were extended to include the effects of
electron-electron interactions by Finkelstein \cite{Finkelstein}, and
later by Castellani {\it et. al.} \cite{CCLM}, who obtained RG
equations for the conductance, as well as the singlet and triplet
interaction coupling constants, starting with a diffusive Fermi liquid
fixed point (for a review, see Ref. \cite{Belitz-Kirkpatrick}). This
seminal work defined a field theoretical language to study
the simultaneous presence of
interactions and disorder. Unfortunately, this approach suffered from
being inconclusive -- since the coupled renormalization group
equations flow to strong coupling, away from the perturbative starting
point of a diffusive Fermi liquid state -- and technically
quite involved. A recent discussion of
Finkelstein's replica theory can
be found in Ref. \cite{Pruisken}.

The recent experimental discovery by Kravchenko and co-workers
\cite{Kravchenko} of a MIT in a two-dimensional Si-MOSFET and the
subsequent discovery of such a transition in other two-dimensional
\cite{Popovic,GaAs,SiGe} electron gas systems has rekindled interest
in sharpening our understanding of the combined effects of disorder
and interactions. In this paper we will rederive the RG equations for
all marginal perturbations of the diffusive Fermi
liquid fixed point (there are no relevant perturbations).
We will do so using the dynamical
Schwinger-Keldysh approach. Our results agree with those obtained
using the replica method \cite{Finkelstein} and disorder-averaged
perturbation theory \cite{CCLM}. As we will show, the
Schwinger-Keldysh and the replica approaches to studying the diffusive
fermi liquid exhibit a very similar structure. It is important to note
that, in the replica solution, there are no subtleties involved the
$N\to 0$ replica limit at the perturbative level, which reflects the
fact that that there is no replica symmetry breaking in the diffusive
fermi liquid state. This is the underlying reason why the
Schwinger-Keldysh and replica solutions, as we will see, look very
much alike, with the dynamical thermal indices for time-ordered and
anti-time-ordered fields behaving as simple bookkeeping devices, just
as in the replica solution. On the other hand, glassy systems which
exhibit replica symmetry breaking, such as the recently proposed
Wigner glass phase \cite{UCLA,Giam} should be rather different and more
interesting.  We will explore these directions in the future. In the
present paper, we will limit ourselves to the study of the diffusive
fermi liquid, which we present below.

We present the Schwinger-Keldysh approach for disordered and
interacting electronic systems in section \ref{sec:Schwinger-Keldysh},
where we show that instead of $N\to 0$ replicas, we only need to
consider $N=2$ species of fields. Using this dynamic approach, we
derive in Section \ref{sec:sigmamodel} a non-linear $\sigma$-model for
interacting diffusion modes in the Schwinger-Keldysh
formalism. For simplicity, we restrict attention to the `unitary' case
considered in Finkelstein's original paper \cite{Finkelstein}: we
assume that the electrons interact through a short-range interaction
and that they are in a magnetic field which affects their orbital
motion, but that the Zeeman coupling vanishes (this situation might be
realizable in GaAs-AlGaAs systems).  By straightforward extension of
the methods used here, we can also treat the more interesting, but
more complicated, `orthogonal' case in which the magnetic field is
turned off.  In a similar vein, we avoid the straightforward but more
involved extension to long-range Coulomb interactions. This
would include applications of our approach
to the quantum Hall plateau
transitions. We emphasize
the similarities and differences with the replicated
$\sigma$-model. In section \ref{sec:param}, we find the diffusive
saddle point and bring the $\sigma$-model into a form which is
convenient for perturbation theory about this saddle point.  This form
is used to derive the Feynman rules, given in section
\ref{sec:frules}, which are needed for one-loop perturbative
calculations.  In section \ref{sec:renorm}, we use these Feynman rules
to discuss the one-loop renormalization of this $\sigma$-model. One of
our main purposes here is pedagogy, so we discuss the diagrammatics in
detail, paying attention to the symmetry factors and minus signs which
prove to be crucial in determining the physics of the model. As we
show, the apparent complexity -- due to the large number of diagrams
-- can be offset by a systematic enumeration. Finally, in sections
\ref{sec:reg} and \ref{sec:disc}, we discuss the resulting RG
equations, a constraint on them following from the Ward identity for
charge conservation, and their physics.

\section{The dynamical Schwinger-Keldysh approach to disordered and
interacting electrons}
\label{sec:Schwinger-Keldysh}

\begin{figure}
\centerline{\psfig{figure=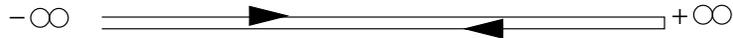,height=.15in,angle=0}}
\vskip 0.5cm
\caption{Keldysh contour, going from $-\infty$ to $\infty$
and back to $-\infty$.}
\label{fig:Keldysh}
\end{figure}

In the Schwinger-Keldysh -- or closed time path -- formalism
\cite{Schwinger,Keldysh,Rammer-Smith}, a functional integral (see also
the formulation by Feynman and Vernon, Ref. \cite{Feynman-Vernon}) is
constructed for the time-evolution of the vacuum state from
$t=-\infty$ to $t=\infty$ and back to $t=-\infty$ (the Keldysh
contour, shown in Fig. \ref{fig:Keldysh}, is just one possible path
for this evolution). Such evolution of the vacuum brings it back to
the initial state, and therefore the vacuum-to-vacuum overlap (or
vacuum persistence $Z$) in the closed path formalism is
1. Consequently, the functional integral is automatically normalized
to $Z=1$ for any realization of the disorder potential, so
disorder-averaged correlation functions can be calculated directly
from the disorder-averaged functional integral. The price which must
be paid is the doubling of the number of the fields in the theory; the
second copy of each field propagates backwards in time.

The application of dynamical approaches, such as the Schwinger-Keldysh
formalism, to disordered systems has been previously proposed. At the
classical level, Martin, Siggia, and Rose \cite{MSR} used a dynamical
approach and explored the independence of the classical generating
functional on the disorder. Sompolinsky \cite{Somp} used this formalism
to study the mean-field theory for the spin glass problem.
Quantum versions of the idea of using
dynamics to enforce $Z=1$ have been proposed by Schuster and Vieira
\cite{Schuster-Vieira} and by Kree \cite{Kree}. However, these
proposals were never stated in a language suitable for calculations in
disordered systems. A concrete formulation was put forward by Horbach
and Sch\"on \cite{Schon}, who developed a time-path formulation for
disordered non-interacting electrons.
Recently, Cugliandolo and Lozano \cite{Leticia} proposed
a closed time path formalism for quantum spin glasses.
In work complementary to ours,
Andreev and Kamenev \cite{anton+alex} use the Schwinger-Keldysh formalism
to address the issue of gauge invariance in disordered
interacting electron systems. They discuss the
single-particle density of states -- which we do not --
but do not derive the full set of coupled RG
equations -- which we do, following Finkelstein.

Here we apply the closed time path formulation, and cast it in a
language appropriate to consider the effect of interactions. We
give an extended discussion of the method in Ref. \cite{CLN1}, where
we discuss, in addition to systems with a natural quantum dynamics, a
non-trivial extention of the method to systems with no natural quantum
or classical dynamics, such as disordered, interacting
statistical field theories. Below we give a short summary of the method.

The Keldysh contour in Fig. \ref{fig:Keldysh} is just one example of a
possible contour for which we can achieve our desired objective,
namely the absence of a denominator in the correlation functions ({\it
i.e.}, $Z=1$). Here we will discuss a more general class of contours originally
devised for the study of field theories at finite temperatures
\cite{Semenoff-Umezawa,Niemi-Semenoff,Landsman-Weert}. These
formulations are carried out in real-time, and their objective was to
circumvent pitfalls from the strong assumptions about analytical
continuation of imaginary-time (such as Matsubara's) formulations.

\begin{figure}
\centerline{\psfig{figure=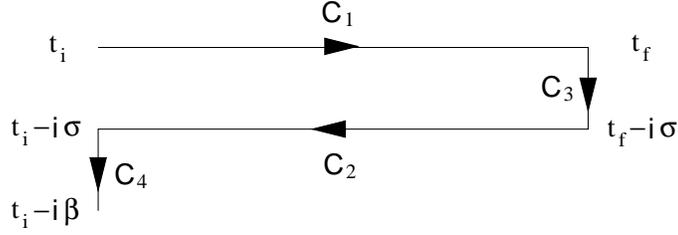,height=1.2in,angle=0}}
\vskip 0.5cm
\caption{Real-time contour separated into four parts that factorize 
into separate contributions: $C_1\cup C_2$ and $C_3\cup C_4$.}
\label{fig:contour}
\end{figure}

At finite temperature, the contour of figure \ref{fig:Keldysh}
is replaced by a contour running from $t=-\infty$ to $t=\infty$
and thence to $t=-\infty-i\beta$. A contour of this type is
displayed in figure \ref{fig:contour}. For such a contour,
there is an important factorization property\cite{CLN1}, shown in
Ref. \cite{Landsman-Weert}, for the contributions from each piece of
the contour to the functional integral $Z$:
\begin{equation}
Z=Z^{C_1\cup C_2}\;Z^{C_3\cup C_4}\ .
\end{equation}
Only $C_1$ and $C_2$ are important in obtaining physical
correlation functions.\cite{factorization}
As a consequence of the
factorization property, we can obtain these correlation functions
from $Z^{C_1\cup C_2}$ which satisfies the normalization
property:
\begin{equation}
{Z^{C_1\cup C_2}} = 1
\end{equation}
Hence, even at finite-temperature, we can work with a
partition function normalized to $1$.
Denoting the fields on the upper ($C_1$) and the lower
($C_2$) pieces of the countour by
\begin{equation}
\phi_1(t) =\phi(t),
\qquad
\phi_2(t) = \phi(t-i\sigma),
\qquad \qquad (t={\rm real})
\end{equation}
one defines a matrix propagator
\begin{equation}
<T_c[\phi_a(t,x)\phi_b^{\dagger}(t',x')]>
=-i\Delta_{ab}(t-t',x-x')
\end{equation}
where $T_c$ denotes ordering of fields according their position along
the contour of Fig. \ref{fig:contour}. The particular choice
$\sigma=\beta/2$ makes the form of the propagators especially
simple. Let $\phi$ denote a complex field, bosonic or fermionic.
The propagator $i\Delta^{ab}$
can be written in terms of the zero-temperature propagator
$i\Delta_0^{ab}$ as
\begin{equation}
i\Delta(\omega,k)=
u(\omega)\; i\Delta_0(\omega,k)\; u^\dagger(\omega)\ ,
\end{equation}
where
\begin{equation}
i\Delta_0(\omega,k)=
\left(\matrix{iG_0(\omega,k) & 0 \cr
                0 & -iG_0^*(\omega,k)}\right)\ ,
\end{equation}
with $iG_0(\omega,k)$ the usual time-ordered propagator
\begin{equation}
iG_0(t-t',x-x')=\langle T[\phi(t,x)\phi^{\dagger}(t',x')]\rangle
\end{equation}
and $-iG^*_0(\omega,k)$, consequently, the anti-time-ordered one
\begin{equation}
-iG^*_0(t-t',x-x')=
\big(\langle T[\phi(t,x)\phi^{\dagger}(t',x')]\rangle\big)^*
=\langle {\bar T}[\phi(t',x')\phi^{\dagger}(t,x)]\rangle
\end{equation}

The matrix $u$ contains the information about the temperature.
For bosonic fields, this matrix is given by
\begin{equation}
u(\omega) = u_B(\omega)=
\left(\matrix{\cosh\Delta\theta_\omega & \sinh\Delta\theta_\omega \cr
                 \sinh\Delta\theta_\omega & \cosh\Delta\theta_\omega }\right)\ ,
\quad 
 {\rm where}  \ \ 
\Delta\theta_\omega=\theta_\omega^T-\theta_\omega^{T=0},
\qquad {\rm and} \qquad
 \cosh^2\theta_\omega^T=\frac{1}{1-e^{-\omega/T}}\ .
\end{equation}
For a fermion field, the matrix is altered to account for the
fermionic statistics, and we have
\begin{equation}
u(\omega) = u_F(\omega)=
\left(\matrix{\cos\Delta\theta_\omega & \sin\Delta\theta_\omega \cr
                 -\Delta\sin\theta_\omega & \cos\Delta\theta_\omega }\right)\ ,
\quad
 {\rm where}  \ \
\Delta\theta_\omega=\theta_\omega^T-\theta_\omega^{T=0},
 \quad {\rm and} \quad
\cos^2\theta_\omega^T=\frac{1}{1+e^{-(\omega-\mu)/T}}\ .
\end{equation}
Notice that at zero temperature $u_{B,F}={\bf 1}$.

All correlation functions in a theory with propagator $\Delta_{ab}$
and interaction ${\cal L}[\phi,\phi^*]$ are obtained  using
the following  functional integral
over two fields $\phi_{1,2}$:
\begin{eqnarray}
Z =\int D\phi_1  D\phi_2\
\exp\left[i\int\left(\frac{1}{2}\phi^*_a[\Delta^{-1}]^{ab} \phi_b
+{\cal L}_I[\phi_1,\phi_1^*]-{\cal L}_I[\phi_2,\phi_2^*]\right) \right]\ ,
\label{eq:T-loop-path-int}
\end{eqnarray}
where all time integrals are to be done from $-\infty$ to $\infty$.
This gives rise to a crucial
relative sign in how the interaction comes in for the
forward propagating field $\phi_1$ and the backward propagating one
$\phi_2$.

\section{Non-linear sigma model within the Schwinger-Keldysh formalism}
\label{sec:sigmamodel}


We start by deriving a non-linear sigma model in the language of the
dynamical Schwinger-Keldysh formalism. The steps are quite similar to the deriva
tion in the
replica formulation.

As shown above, we formulate the problem in terms of a real time path
integral
\begin{equation}
Z=\int D\psi^\dagger D\psi \; e^{i S[\psi^\dagger,\psi]}
\end{equation}
for a fermionic field
\[
\psi = \left(\matrix{
\psi_1 \cr
\psi_2
}
\right)
\]
comprised of two components $\psi_i$ labelled by the thermal indices
$i=1,2$ (in addition to other indices such as spin). The index $i=1$
corresponds to time-ordered fields, and $i=2$ to anti-time-ordered
ones.

\subsection{The free action}
The free part of the action for these fermions can be written as
\begin{equation}
S_0[\psi^\dagger,\psi]=
\int d^d x \int \frac{d\omega}{2\pi} \ \ \psi^\dagger(x,\omega)\;
u_F(\omega) \; \sigma_3 \; 
\left[\matrix{
\omega + \frac{\nabla^2}{2m} + \epsilon_F + i\eta\; {\rm sgn} \omega & 0 \cr
0 &\omega  + \frac{\nabla^2}{2m} + \epsilon_F - i\eta\; {\rm sgn}\omega
}
\right] u^\dagger_F(\omega) \;\psi(x,\omega) \ ,
\end{equation}
from which we obtain the free matrix propagator
\begin{equation}
\widehat{\psi \psi^\dagger}=
i\Delta(\omega,k)=
u_F(\omega)\; i\Delta_0(\omega,k)\; u^\dagger_F(\omega)\ ,
\end{equation}
where
\begin{equation}
i\Delta_0(\omega,k)=
\left[\matrix{iG_0(\omega,k) & 0 \cr
                0 & -iG^*_0(\omega,k)}\right]\ ,
\end{equation}
and $iG_0$ is the time-ordered propagator
\[
iG_0(\omega,k)=\frac{i}{\omega - \frac{k^2}{2m} + \epsilon_F 
+ i \eta\; {\rm sgn} \omega}
\ .
\]
Notice that the anti-time-ordered propagator is the adjoint of
$iG_0$. The matrix $u_F(\omega)$ containing the temperature
dependence was defined above.
In particular, $u_F(\omega)={\bf 1}_{\rm thermal}$ at $T=0$. In this
case we have
\begin{equation}
S_0[\psi^\dagger,\psi]=
\int d^d x \int \frac{d\omega}{2\pi}\ \ \bar\psi(x,\omega)\;
\left[\left(\omega + \frac{\nabla^2}{2m} + \epsilon_F \right) {\bf 1}
+ i\eta\;\sigma_3\; {\rm sgn} \omega \right]
\;\psi(x,\omega) \ ,
\end{equation}
where in the last line we defined $\bar\psi=\psi^\dagger \sigma_3$.

\subsection{Disorder term}

The static disorder contribution to the action is
\begin{equation}
S_{V}[\psi^\dagger,\psi]=
\int d^d x \int \frac{d\omega}{2\pi} \ V(x) \ \psi^\dagger(x,\omega)\;
\sigma_3  \;\psi(x,\omega) \ \ .
\end{equation}
Notice that any interaction term entering the action has the following
properties:
\begin{enumerate}
\item
It does not mix the two fields $\psi_{1,2}$. There will be mixing
after disorder averaging, but not before.
\item
The contributions from the two thermal components $i=1,2$ enter with
opposite signs (thus the $\sigma_3$) due to the negative sign for the
time integration along the anti-time-ordered branch.
\end{enumerate}
The disorder potential is assumed to be Gaussian distributed according
to
\begin{equation}
P[V(x)]\propto
e^{-\frac{1}{2u} \int dx \;V^2(x)}\ .
\end{equation}
We then integrate out the disorder, using the fact that the 
Schwinger-Keldysh 
formulation cancels the denominator problem:
\begin{equation}
e^{iS_{\rm rand}[\psi^\dagger,\psi]}=
\int DV \; P[V]\; e^{i S_{V}[\psi^\dagger,\psi]}\ ,
\end{equation}
thus generating a four fermion term
\begin{eqnarray}
S_{\rm rand}[\psi^\dagger,\psi]&=&i\;\frac{u}{2}\;
\int d^d x \int \frac{d\omega}{2\pi} \frac{d\omega'}{2\pi}
\;\psi^\dagger(x,\omega)\;
\sigma_3  \;\psi(x,\omega) \ \psi^\dagger(x,\omega')\;
\sigma_3  \;\psi(x,\omega') \\
&=&i\;\frac{u}{2}\;
\int d^d x \int \frac{d\omega}{2\pi} \frac{d\omega'}{2\pi}
\;\bar\psi(x,\omega)\;
\psi(x,\omega) \ \bar\psi(x,\omega')  \;\psi(x,\omega')\ .\nonumber
\end{eqnarray}

Let us now introduce the Hubbard-Stratonovich matrix field $Q$ that
decouples the four fermions (the matrix $Q$ is chosen to be
Hermitian). We will choose the decoupling in the diffusive
particle-hole channel. As in the case of the original diffusive Fermi
liquid study by Finkelstein, here we will not consider the cooperon
contributions. We have
\begin{equation}
e^{iS_{\rm rand}[\bar\psi,\psi]}=
\int DQ \;\;
e^{-\frac{1}{2u}
\int d^d x \;{\rm tr}\; Q^2}\;
e^{iS_{\rm HS}[Q,\bar\psi,\psi]}
\end{equation}
where
\begin{equation}
S_{\rm HS}[Q,\bar\psi,\psi]=i
\int d^d x \int \frac{d\omega}{2\pi} \frac{d\omega'}{2\pi}\;
\bar\psi(x,\omega)\;Q_{\omega\omega'}(x)\;\psi(x,\omega')\ .
\end{equation}
The matrix $Q$ has indices in three separate spaces, {\it i.e.}, it
is assembled as a direct product in energy, thermal, and spin
spaces:
\[
Q^{{ij},{\alpha\beta}}_{\omega\omega'}
= \left[Q^{{ji},{\beta\alpha}}_{\omega'\omega}\right]^*
\quad\quad{\rm hermitian}.
\]
Thus, the trace of $Q^2$ corresponds to
\[
{\rm tr}\;Q^2=\int \frac{d\omega}{2\pi} \frac{d\omega'}{2\pi}\;
\sum_{i,j=1}^2\;\sum_{\alpha,\beta=1}^2\;
Q^{{ij},{\alpha\beta}}_{\omega\omega'}\;
Q^{{ji},{\beta\alpha}}_{\omega'\omega}
\]
When we write $Q_{\omega\omega'}$ we mean a matrix whose elements are
matrices in thermal ($i,j=1,2$) and spin spaces ($\alpha,\beta=1,2$).

We can then write an expression for $Z$ as a functional integral over
$\bar\psi,\psi$ and $Q$:
\begin{equation}
Z=\int DQ \int D\bar\psi D\psi
\;
e^{-\frac{1}{2u}
\int d^d x \;{\rm tr}\; Q^2}\;\;
e^{iS_0[\bar\psi,\psi]}\;
e^{iS_{\rm HS}[Q,\bar\psi,\psi]}
\end{equation}
where, writing $S_0$ and $S_{\rm HS}$ explicitly in terms of sums over
the energy, thermal and spin indices
\begin{equation}
S_0+
S_{\rm HS}
=\int d^d x \int  \frac{d\omega}{2\pi} \frac{d\omega'}{2\pi}\;
\sum_{ij,\alpha\beta}\;
\bar\psi_{i,\alpha}(x,\omega)\;\left\{
\left[\left(\omega + \frac{\nabla^2}{2m}+\epsilon_F \right) \delta_{ij}
+ i\eta\;\sigma^{ij}_3\; {\rm sgn} \omega \right]\;\delta_{\alpha\beta}\;
\delta_{\omega\omega'}
+i Q^{{ij},{\alpha\beta}}_{\omega\omega'}(x)\right\}
\;\psi_{j,\beta}(x,\omega')
\ .
\end{equation}
Upon integrating out the fermions we obtain an effective theory for
the matrix $Q$ field,
\begin{equation}
Z=\int DQ \;
e^{-\frac{1}{2u}
\int d^d x \;{\rm tr}\; Q^2}\;
e^{\int d^d x \;
{\rm tr}\ln\;
\left[-i \Omega\; -i \left(\frac{\nabla^2}{2m}+\epsilon_F\right)
\;{\bf 1}+\eta \Lambda +Q(x)\right]}
\ ,
\label{eq:sigma-no-int}
\end{equation}
where we introduced $\Omega$ and $\Lambda$ as matrices in the energy
$\otimes$ thermal $\otimes$ spin spaces:
\begin{equation}
\Omega^{ij,\alpha\beta}_{\omega\omega'}=\omega\;\delta_{\omega\omega'}
\;\delta_{ij}\;\delta_{\alpha\beta}
\end{equation}
and
\begin{equation}
\Lambda^{ij,\alpha\beta}_{\omega\omega'}={\rm sgn}\;\omega
\;\delta_{\omega\omega'}
\;\sigma_3^{ij}\;\delta_{\alpha\beta}
\qquad
{\rm or}
\qquad
\Lambda=\underbrace{\Sigma_3}_{\rm Energy} \otimes
\underbrace{\sigma_3}_{\rm Thermal} \otimes
\underbrace{\bf 1}_{\rm Spin}
\label{eq:lambda}
\end{equation}
with $\Sigma_3$ defined as matrix in energy space that is diagonal and
whose entries are $+1$ for positive energies and $-1$ for negative
ones.

The expression Eq. (\ref{eq:sigma-no-int}) is the non-linear sigma
model for a non-interacting system. The information necessary to
dispose of disconnected loops generated by the disorder average is
encoded in the matrix structure for $\Omega$ and $\Lambda$. This
structure ensures that we can work with only two indices, $i,j=1,2$,
instead of $N$ replicas with $N\to 0$ at the end. Notice, however,
that very much like in the replicated sigma model, the disorder
connects $Q$ components sitting on different indices. Next, we will
add the contributions due to interactions.

\subsection{Interaction term}
Let us consider the case of short range interactions, in the singlet
and triplet channels:
\[
S_{\rm int}=S_1+S_2
\]
where
\begin{eqnarray}
S_1[\psi^\dagger,\psi]&=&{\pi\Gamma_1}
\int d^d x \int \frac{d\omega}{2\pi} \frac{d\omega'}{2\pi}\;
\frac{d\epsilon}{2\pi}
\;\sum_{ij,\alpha\beta}\;
\psi_{i,\alpha}^\dagger(x,\omega)\;
\psi_{i,\beta}(x,\omega+\epsilon)
\; \sigma^{ij}_3  \;\psi_{j,\beta}^\dagger(x,\omega'+\epsilon)\;
\psi_{j,\alpha}(x,\omega') \\
&=& {\pi\Gamma_1}
\int d^d x \int \frac{d\omega}{2\pi} \frac{d\omega'}{2\pi}\;
\frac{d\epsilon}{2\pi}
\;\sum_{ij,\alpha\beta}\;
\bar\psi_{i,\alpha}(x,\omega)\;
\psi_{i,\beta}(x,\omega+\epsilon)
\; \sigma^{ij}_3  \;\bar\psi_{j,\beta}(x,\omega'+\epsilon)\;
\psi_{j,\alpha}(x,\omega') \nonumber
\end{eqnarray}
and
\begin{eqnarray}
S_2[\psi^\dagger,\psi]&=&- {\pi\Gamma_2}
\int d^d x \int \frac{d\omega}{2\pi} \frac{d\omega'}{2\pi}
\frac{d\epsilon}{2\pi} \;\sum_{ij,\alpha\beta}\;
\psi_{i,\alpha}^\dagger(x,\omega)\;
\psi_{i,\alpha}(x,\omega+\epsilon)
\; \sigma^{ij}_3  \;\psi_{j,\beta}^\dagger(x,\omega'+\epsilon)\;
\psi_{j,\beta}(x,\omega') \\
&=&-{\pi\Gamma_2}
\int d^d x \int 
\frac{d\omega}{2\pi} \frac{d\omega'}{2\pi}
\frac{d\epsilon}{2\pi} \;\sum_{ij,\alpha\beta}\;
\bar\psi_{i,\alpha}(x,\omega)\;
\psi_{i,\alpha}(x,\omega+\epsilon)
\; \sigma^{ij}_3  \;\bar\psi_{j,\beta}(x,\omega'+\epsilon)\;
\psi_{j,\beta}(x,\omega') \nonumber
\end{eqnarray}
Notice that
\begin{enumerate}
\item
Once again, the interaction does not mix the fields $\psi$ in
different thermal indices $i\ne j$; this is enforced by the $\sigma_3$
matrix in thermal space.
\item
The contributions from the two thermal components $i=1,2$ enter with
opposite signs due to the negative sign for the time integration along
the anti-time-ordered branch.
\item
Because all four fermions sit on the same index $i$, using the
definition $\bar\psi_{j,\alpha}=\psi^\dagger_{i,\alpha}\sigma^{ij}_3$
does not change the structure of the interaction as the positive or
negative signs from the $\sigma_3$ matrices always come in pairs.
\end{enumerate}

We now introduce two Hubbard-Stratonovich fields $X$ and $Y$ to
decouple the four fermion interactions.
\begin{eqnarray}
e^{iS_1[\bar\psi,\psi]}&=&
\int DY \;
e^{iS_{\rm y}[Y]}\;
e^{i\sqrt{2\pi\Gamma_1}\;\int d^d x \int 
\frac{d\omega}{2\pi} \frac{d\omega'}{2\pi} \;
\sum_{ij,\alpha\beta}\;
\bar\psi_{i,\alpha}(x,\omega)
\;Y^{{ij},{\alpha\beta}}_{\omega\omega'}(x)
\;\psi_{j,\beta}(x,\omega')}
\label{eq:Sx}\\
e^{iS_2[\bar\psi,\psi]}&=&
\int DX \;
e^{iS_{\rm x}[X]}\;
e^{\sqrt{2\pi\Gamma_2}\;\int d^d x \int 
\frac{d\omega}{2\pi} \frac{d\omega'}{2\pi}\;
\sum_{ij,\alpha\beta}\;
\bar\psi_{i,\alpha}(x,\omega)
\;X^{{ij},{\alpha\beta}}_{\omega\omega'}(x)
\;\psi_{j,\beta}(x,\omega')}
\label{eq:Sy}
\end{eqnarray}
where
\[
X^{{ij},{\alpha\beta}}_{\omega\omega'}=
{\rm X}^{i,{\alpha\beta}}(\omega-\omega')\;\delta_{ij}
\qquad
Y^{{ij},{\alpha\beta}}_{\omega\omega'}=
{\rm Y}^{i}(\omega-\omega')\;\delta_{\alpha\beta}\;\delta_{ij}
\]
Notice that the matrices $X^{{ij},{\alpha\beta}}_{\omega\omega'}$ and
$Y^{{ij},{\alpha\beta}}_{\omega\omega'}$ depend only on the energy
difference $\omega-\omega'$. We choose $X$ Hermitian ($X=X^\dagger$),
and $Y$ to be anti-Hermitian ($Y=-Y^\dagger$); we could alternatively,
have absorbed a factor of $i$ into $Y$ and made it Hermitian. The
action for the matrices $X$ and $Y$ is
\begin{eqnarray}
S_{\rm x}[X]&=&\frac{1}{2}\int d^d x \int \frac{d\epsilon}{2\pi} \;
\; {\rm X}^{i,\alpha\beta}(\epsilon)
\;\sigma^{ij}_3
\;{\rm X}^{j,\beta\alpha}(-\epsilon) 
\label{eq:X}
\\
S_{\rm y}[Y]&=&\frac{1}{2}\int d^d x \int \frac{d\epsilon}{2\pi} \;
\; {\rm Y}^{i}(\epsilon)
\;\sigma^{ij}_3
\;{\rm Y}^{j}(-\epsilon)
\label{eq:Y}
\end{eqnarray}
The action for the $i=1,2$ components of $X$ and $Y$ in
Eqs. (\ref{eq:X},\ref{eq:Y}) come with opposite signs, which is
necessary to generate the correct sign in the four fermion terms for
time-ordered and anti-time-ordered pieces. The fact that at least one
of the components is not positive definite is not a problem here
because we work with a real time path integral and have the $i$ factor
in front of the actions $S_{\rm x}$ and $S_{\rm y}$ as in
Eqs. (\ref{eq:Sx},\ref{eq:Sy}).

Including the fields $X$ and $Y$ together with the matrix field $Q$,
and integrating out the fermions, we can write the partition function
(vacuum persistence) as:
\begin{eqnarray}
Z&=&\int DQ \; DX \; DY \;
e^{-\frac{1}{2u}
\int d^d x \;{\rm tr}\; Q^2}\;
e^{iS_{\rm x}[X]}\;
e^{iS_{\rm y}[Y]} \nonumber \\
&\ &\;\;\;\;\;\;\;
e^{\int d^d x \;
{\rm tr}\ln\;
\left[-i \Omega\; -i \left(\frac{\nabla^2}{2m}+\epsilon_F \right)
\;{\bf 1}+\eta \Lambda + Q(x)
-\sqrt{2\pi\Gamma_2}\; X(x)
-i\sqrt{2\pi\Gamma_1}\; Y(x)
\right]}
\ ,
\label{eq:sigma-int}
\ .
\end{eqnarray}

\subsection{Expansion around the diffusive Fermi liquid fixed point}

We will now expand the ${\rm tr}\ln$ aroung the $Q$ saddle of a
diffusive Fermi liquid. The saddle point solution for the
non-interacting problem is $Q=\pi u \nu_0\; \Lambda$, where $\nu_0$ is
the density of states at the fermi level and $\Lambda$ is given in
Eq. (\ref{eq:lambda}). The matrix $Q$ can be rescaled so that the
saddle point can be parametrized as $Q=U^\dagger \Lambda U$. The
matrix $Q$ in the non-linear sigma model then satisfy the constraints
\begin{equation}
Q=Q^\dagger ,
\qquad
Q^2={\bf 1} ,
\qquad
{\rm tr}\; Q =0\ .
\end{equation}
The expansion is achieved by allowing a slow position dependence in
the matrix $U\to U(x)$, and considering small couplings $\Gamma_{1,2}$
and small energies $\omega$. Expanding Eq. (\ref{eq:sigma-int}) around
the saddle point, we have
\begin{eqnarray}
Z&=&\int DX \; DY \;DQ \;
e^{iS_{\rm x}[X]}\;
e^{iS_{\rm y}[Y]} \nonumber \\
&\ &\;\;\;\;\;\;\;
e^{-\frac{1}{2}\int d^d x \;
\left\{D\;
{\rm tr}\;(\nabla Q)^2\;
+4z\;{\rm tr}\; [(i\Omega-\eta\Lambda)\; Q]\;
+2\sqrt{2\pi\Gamma_2}\;{\rm tr}\; (X Q)\;
+2i\sqrt{2\pi\Gamma_1}\;{\rm tr}\; (Y Q)
\right\}}
\ ,
\end{eqnarray}
where we have also absorbed the rescaling of the saddle into the
couplings $\Gamma_{1,2}$, as well as $D$ and $z$.
Integrating out $X$ and $Y$, we obtain $Z=\int
DQ\; e^{-S[Q]}$, with an effective action for $Q$ as follows:
\[
S[Q]=S_D[Q] + S_{e-e}[Q]
\]
where
\begin{eqnarray}
S_D[Q]&=&\frac{1}{2}
\int d^d x \;
\left[D\;
{\rm tr}\;(\nabla Q)^2\;
+4z\;{\rm tr}\; [(i\Omega-\eta\Lambda)\; Q]\right]
\label{eq:Sd}
\\
S_{e-e}[Q]&=&\ \ 
\int d^d x \;
\left[i\pi\Gamma_1 \; Q\gamma_1 Q 
-i\pi\Gamma_2 \; Q\gamma_2 Q \; \right]
\label{eq:Se-e}
\end{eqnarray}
and the contractions $Q\gamma_1 Q$ and $Q\gamma_2 Q$ correspond to
\begin{eqnarray}
Q\gamma_1 Q &=&
\sum_{i,\alpha\beta}
\int \frac{d\omega_1}{2\pi} \frac{d\omega_2}{2\pi} 
\frac{d\omega_3}{2\pi} \frac{d\omega_4}{2\pi} 
\;\;
Q^{{ii},{\alpha\alpha}}_{\omega_1\omega_2}\;
Q^{{ii},{\beta\beta}}_{\omega_3\omega_4}\;\;
\sigma^{ii}_3\;\;
\;2\pi
\delta(\omega_1-\omega_2+\omega_3-\omega_4)
\\
Q\gamma_2 Q &=&
\sum_{i,\alpha\beta}
\int \frac{d\omega_1}{2\pi} \frac{d\omega_2}{2\pi} 
\frac{d\omega_3}{2\pi} \frac{d\omega_4}{2\pi} 
\;\;
Q^{{ii},{\alpha\beta}}_{\omega_1\omega_2}\;
Q^{{ii},{\beta\alpha}}_{\omega_3\omega_4}\;\;
\sigma^{ii}_3\;\;
\;2\pi
\delta(\omega_1-\omega_2+\omega_3-\omega_4)
\end{eqnarray}

At this stage we would like to compare the expressions for the
effective non-linear sigma model
Eqs. (\ref{eq:Sd},{\ref{eq:Se-e}). First, notice that the structure is
very close to Finkelstein's replicated model; however, here we have
only two ``replicas'' corresponding to the two thermal indices. Also, the
interaction terms $Q\gamma_1 Q$ and $Q\gamma_2 Q$ do not couple the
two thermal spaces much in the same way that the interaction does not
couple different replicas. In the Schwinger-Keldysh  approach the two thermal
indices appear with an opposite sign, hence the $\sigma^{ii}_3$, in
contrast to $\delta_{ij}$ as is the case of replicas.

\section{parametrization of the saddle}
\label{sec:param}

We have used above the parametrization $Q=U^\dagger \Lambda U$, where
$\Lambda={\Sigma_3}_{\rm Energy} \otimes {\sigma_3}_{\rm Thermal}
\otimes {\bf 1}_{\rm Spin}$. The tensor product of the ${\Sigma_3}$ in
energy space and $\sigma_3$ in thermal space makes the parametrization
of the saddle rather cumbersome for carrying out calculations. This
can be resolved through the use of a transformation that allows us to
parametrize around the saddle point
\[
\tilde\Lambda^{ij,\alpha\beta}_{\omega\omega'}={\rm sgn}\;\omega
\;\delta_{\omega\omega'}
\;\delta_{ij}\;\delta_{\alpha\beta}
\qquad
{\rm or}
\qquad
\tilde\Lambda=\underbrace{\Sigma_3}_{\rm Energy} \otimes
\underbrace{\bf 1}_{\rm Thermal} \otimes
\underbrace{\bf 1}_{\rm Spin}\ \ .
\]
In this case the thermal and spin structures for the saddle become
trivial, and we need only to worry about the energy structure. This transformati
on is achieved through
\[
Q \to T^\dagger Q T
\]
where
\[
T^{ij,\alpha\beta}_{\omega\omega'}=\cases{
\delta_{\omega,\omega'}\;\delta_{ij}\;\delta_{\alpha\beta}
&\mbox{$i=1$}\cr
\delta_{\omega,-\omega'}\;\delta_{ij}\;\delta_{\alpha\beta}
&\mbox{$i=2$}\cr
}
\]
The matrix $T$ can be viewed as a direct product of matrices in
energy, thermal and spin spaces: $T=U(i) \otimes {\bf 1}_{\rm thermal}
\otimes {\bf 1}_{\rm spin}$, where $U(i)$ is a rotation in energy space that
dependes on the thermal index. This transformation leaves the form of
the action invariant, with the exception of $\Lambda\to\tilde\Lambda$
and $\Omega\to\tilde\Omega$, where
\[
\tilde\Omega^{ij,\alpha\beta}_{\omega\omega'}=
\omega
\;\delta_{\omega\omega'}
\;\sigma_3^{ij}\;\delta_{\alpha\beta}\ \ .
\]
It is easy to check that the forms of $Q\gamma_1 Q$ and $Q\gamma_2 Q$
remain unchanged upon using the transformation $T$ (the interaction
does not mix thermal indices, so for $i=1$ this is trivial, and for
$i=2$ one can absorb all changes in sign by redefining the integration
variables).

The parameterization of $Q$ around the $\tilde\Lambda$ saddle becomes
simple and familiar, as in the work of Hikami \cite{Hikami}:
\begin{equation}
\matrix{
\matrix{\matrix{ \ \ \ \ \ \ \omega'>0 \cr\;\cr} & \ \ \ \ \ \ \
\matrix{ \omega'<0 \cr\;\cr} \cr}
& \; \cr
Q=\left[\matrix{ \matrix{\sqrt{1-V^\dagger V}\cr\;\cr} &
\matrix{ V^\dagger\cr\;\cr}  \cr
V& -\sqrt{1-V V^\dagger } \cr}\right]
& \matrix{ \matrix{ \omega>0 \cr\;\cr} \cr
\omega<0 \cr} \cr
}
\label{eq:Q-V}
\end{equation}
where the matrix $V_{\omega\omega'}$ has row indices with negative
energies ($\omega<0$) and column indices with positive energies
($\omega'>0$). As in several other works (see, for example, the work
by McKane and Stone \cite{Mike}, and by Grilli and Sorella
\cite{Grilli&Sorella}), we will proceed with the study of the
nonlinear sigma model by expanding in powers of the $V,V^\dagger$
matrices:
\[
Q=\sum_n Q^{(n)}
\]
where $n$ is the order in the power series expansion in
$V,V^\dagger$. For example,
\[
Q^{(0)}=
\left[\matrix{1&0\cr 0&-1\cr} \right]
\ ,
\quad
Q^{(1)}=
\left[\matrix{0&V^\dagger\cr V& 0\cr} \right]
\ ,
\quad
Q^{(2)}=
\left[\matrix{-\frac{1}{2}V^\dagger V& 0\cr
0&\frac{1}{2}V V^\dagger \cr} \right]
\]
\[
Q^{(3)}=0
\ ,
\quad
Q^{(4)}=
\left[\matrix{-\frac{1}{8}(V^\dagger V)^2 &
0\cr 0&\frac{1}{8}(V V^\dagger)^2 \cr} \right]
\ ,
\quad
\dots
\]

The quadratic piece of the expansion of the action $S_D[Q]$ in powers
of $V,V^\dagger$ is
\begin{eqnarray}
S^{(2)}_D[Q]&=&\frac{1}{2}
\int d^d x \;
\left[D\;
{\rm tr}\;(\nabla Q^{(1)})^2\;
+4z\;{\rm tr}\; [(i\tilde\Omega-\eta\tilde\Lambda)\; Q^{(2)}]\right]
\\
&=&
\frac{1}{2}
\int d^d x \;
\int \frac{d\omega}{2\pi}
\int \frac{d\omega'}{2\pi}
\sum_{ij,\alpha\beta}
\left[D\;
\nabla {Q^{(1)}}^{ij,\alpha\beta}_{\omega\omega'}
\nabla {Q^{(1)}}^{ji,\beta\alpha}_{\omega'\omega}
+\; 4z\;(i\tilde\Omega^{ij,\alpha\beta}_{\omega\omega'}
-\eta\tilde\Lambda^{ij,\alpha\beta}_{\omega\omega'})
\; {Q^{(2)}}^{ji,\beta\alpha}_{\omega'\omega}
\right]
\\
&=&
\int d^d k \;
\int_{\omega>0} \frac{d\omega}{2\pi}
\int_{\omega'<0} \frac{d\omega'}{2\pi}
\sum_{ij,\alpha\beta}
\left[D k^2-iz\;(\omega\;\sigma^{ii}_3-\omega'\;\sigma^{jj}_3)\right]\;
\;
{V^\dagger}^{ij,\alpha\beta}_{\omega\omega'}(k)\;
{V}^{ji,\beta\alpha}_{\omega'\omega}(-k)
\end{eqnarray}
and
\begin{equation}
S^{(2)}_{e-e}[Q]=
\int d^d x \;
\left[i\pi\Gamma_1 \; Q^{(1)}\gamma_1 Q^{(1)} 
-i \pi\Gamma_2 \; Q^{(1)}\gamma_2 Q^{(1)} \; \right]
\end{equation}
with
\begin{eqnarray}
\label{gammastructure}
Q^{(1)}\gamma_1 Q^{(1)} &=&
\ \ \sum_{i,\alpha\beta}
\int \frac{d\omega_1}{2\pi} \frac{d\omega_2}{2\pi} 
\frac{d\omega_3}{2\pi} \frac{d\omega_4}{2\pi}\;\;
{Q^{(1)}}^{{ii},{\alpha\alpha}}_{\omega_1\omega_2}\;
{Q^{(1)}}^{{ii},{\beta\beta}}_{\omega_3\omega_4}\;\;
\sigma^{ii}_3\;\;
2\pi\delta(\omega_1-\omega_2+\omega_3-\omega_4)
\\
&=&
2\sum_{i,\alpha\beta}
\int_{\omega_1,\omega_4 >0}
\frac{d\omega_1}{2\pi} \frac{d\omega_4}{2\pi}\;
\int_{\omega_2,\omega_3 <0}
\frac{d\omega_2}{2\pi} \frac{d\omega_3}{2\pi}\;\;
{V^\dagger}^{{ii},{\alpha\alpha}}_{\omega_1\omega_2}\;
V^{{ii},{\beta\beta}}_{\omega_3\omega_4}\;\;
\sigma^{ii}_3\;\;
2\pi\delta(\omega_1-\omega_2+\omega_3-\omega_4)
\\
&\ & \nonumber
\\
Q^{(1)}\gamma_2 Q^{(1)} &=&
\ \ \sum_{i,\alpha\beta}
\frac{d\omega_1}{2\pi} \frac{d\omega_2}{2\pi} 
\frac{d\omega_3}{2\pi} \frac{d\omega_4}{2\pi}\;\;
{Q^{(1)}}^{{ii},{\alpha\beta}}_{\omega_1\omega_2}\;
{Q^{(1)}}^{{ii},{\beta\alpha}}_{\omega_3\omega_4}\;\;
\sigma^{ii}_3\;\;
2\pi\delta(\omega_1-\omega_2+\omega_3-\omega_4)
\\
&=&
2\sum_{i,\alpha\beta}
\int_{\omega_1,\omega_4 >0}
\frac{d\omega_1}{2\pi} \frac{d\omega_4}{2\pi}\;
\int_{\omega_2,\omega_3 <0}
\frac{d\omega_2}{2\pi} \frac{d\omega_3}{2\pi}\;\;
{V^\dagger}^{{ii},{\alpha\beta}}_{\omega_1\omega_2}\;
V^{{ii},{\beta\alpha}}_{\omega_3\omega_4}\;\;
\sigma^{ii}_3\;\;
2\pi\delta(\omega_1-\omega_2+\omega_3-\omega_4)
\end{eqnarray}
Notice that the contribution to quadratic order coming from
$Q^{(2)}\gamma Q^{(0)}$ vanishes due to the trace, as do contributions
to any order in the expansion of $Q\gamma Q$ where $Q^{(0)}$ appears
as one of the two $Q$ terms.

\section{Feynman Rules}
\label{sec:frules}

We now present the Feynman rules which follow
from the action ${S_D} + {S_{e-e}}$.
Our theory is parametrized most succinctly
in terms of the $V$ matrix used above.
The basic propagator is the $\langle VV\rangle$
propagator or diffusion propagator
which follows from $S_D$. It is represented by a
thick line, as in figure \ref{feynrules}(a),
but it can be understood as a double-line
comprised of an electron and a hole.
It carries momentum $q$ and the two
frequencies, $\epsilon$, $\epsilon+\omega$ of the
electron and hole. The arrow is in the direction
of the larger frequency. The propagator also carries
two thermal indices and two
spin indices (one each for the electron and the hole).
The action ${S_D}+{S_{e-e}}$ contains terms
of all orders in $V$. However, not all of these
are needed at the one-loop level, as we now show.

Let us start by counting which vertices we need to include in the
renormalization program. We can determine the loop order of any
diagram \cite{Grilli&Sorella,Brezin} by considering the number $n_v$
of vertices with $v$ legs, and the number of external lines $E$. The
number of legs in all vertices must equal the number of external legs
plus twice the number of internal lines (each internal line belongs to
two vertices):
\begin{equation}
E+ 2I =\sum_{v} v \; n_v \ .
\end{equation}
The loop order is the net number of momentum integrals, which is the
number of internal lines minus the number of contraints from momentum
conservation in each vertex, plus an overall momentum conservation
constraint
\begin{equation}
L= I - \sum_{v}  n_v + 1\ .
\end{equation}
Thus,
\begin{equation}
L= \sum_{v} n_v \; \left(\frac{1}{2} v-1 \right) -\frac{1}{2} E + 1\ .
\end{equation}

With these relations in hand, it is simple to check that we only need
to consider three and four point vertices in the one-loop ($L=1$)
renormalization of the sigma model. As mentioned above, the RG flows
can be obtained from the two-point function alone, so we set $E=2$.
Thus, we have
\begin{equation}
1=\sum_{v} n_v \; \left(\frac{1}{2} v-1 \right) 
\end{equation}
which has only two solutions: ($v=3$, $n_v=2$) or ($v=4$,
$n_v=1$).

Hence, at the one-loop level, we need only consider the two-point
vertex which follows from the expansion of $S_{e-e}$
to quadratic order in the $V$ matrix; the triangle vertices
which follow from its expansion to third order in $V$; and
the square vertices which follow from its expansion to
fourth order as well as the expansion of $S_D$ to
fourth order.
The index structure of the former follows from
(\ref{gammastructure}). The index structure of the latter two
follow from the fact that the thick lines
are understood as splitting at the vertices
to give the thin lines
of the vertices; frequencies, spin, and
thermal indices from the thick lines then split and
follow the thin lines, as we demonstrate with
the spin indices in figure \ref{feynrules}(c). At the interaction `dots',
they follow the same rules as at the two-point
vertices. In figure \ref{feynrules}(b), there is a matrix structure
${\delta^{\alpha\gamma}}{\delta^{\mu\nu}}$
if the interaction `dot' is a $\Gamma_1$
and ${\delta^{\alpha\mu}}{\delta^{\gamma\nu}}$
if the interaction `dot' is a $\Gamma_2$.
This is abbreviated as $\Gamma$.
The frequencies have an overall $\delta$-function
at the `dot'. The Feynman rules are summarized below.

\begin{figure}
\centerline{\psfig{figure=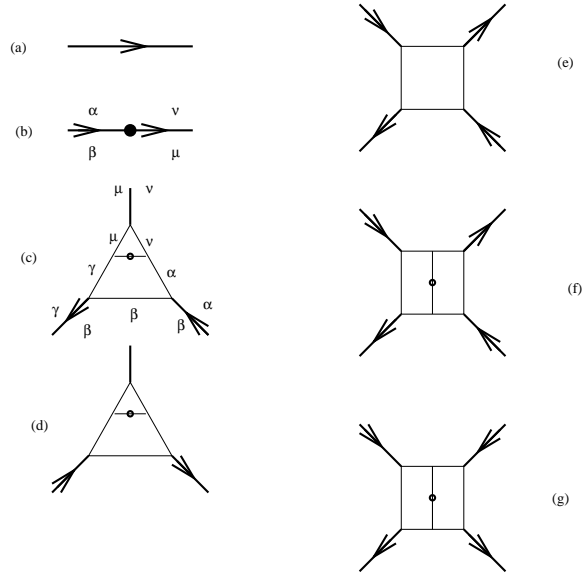,height=3.0in,angle=-90}}
\vskip 0.5cm
\caption{The propagator and two-, three-, and four-point
vertices.}
\label{feynrules}
\end{figure}

{\bf Propagator and Diffusion Vertex.} At the one-loop
level, we need only consider the first two terms in the
expansion of $S_D$. They are the propagator and
the diffusion vertex. The propagator is given by:
\begin{equation}
\label{prop}
{\rm (a)} \,\, = \,\,
\langle {V^\dagger}^{ij,\alpha\beta}_{{\omega_1}{\omega_2}}(q)\;
{V}^{kl,\mu\nu}_{{\omega_3}{\omega_4}}(-q) \rangle
\,\, =\,\,
\frac{1}{-iz\left({\omega_1}{\sigma^{ii}_3} - 
{\omega_2}{\sigma^{jj}_3}\right) + D{q^2}}
\;{\delta^{il}}{\delta^{jk}}\;\;
{\delta^{\alpha\nu}}{\delta^{\beta\mu}}\;\;
2\pi\delta\left({{\omega_1}-{\omega_4}}\right)\;
2\pi\delta\left({{\omega_2}-{\omega_3}}\right)
\end{equation}

The second term in the expansion of $S_D$
is the box, (e). It is given by:
\begin{equation}
{\rm (e)} \,\, = \frac{D}{8}\,\left[2\left({q_1}\cdot{q_3}+
{q_2}\cdot{q_4}\right)+\left({q_1}+{q_3}\right)\cdot
\left({q_2}+{q_4}\right)\right] 
+ i\frac{z}{16}\left[{\omega_1}{\sigma^{ii}_3} - 
{\omega_2}{\sigma^{jj}_3} + {\rm perm.}\right]\ .
\end{equation}
We have omitted in the expression above the momentum and frequency conserving
delta functions. The $q_i$'s are the momenta on the four external legs.
These legs can have arrows alternating between in and out as one goes around
the box, with two pointing in and two pointing out.

{\bf Interaction Vertices.}
The two-point vertex results from the lowest order term in the
expansion of $S_{e-e}$:
\begin{equation}
{\rm (b)} \,\, = \,\,2\pi i\left(
-{\Gamma_1}\,{\delta^{\alpha\beta}}
{\delta^{\mu\nu}} + {\Gamma_2}\,
{\delta^{\alpha\nu}}{\delta^{\beta\mu}}\right)\,
{\sigma_3^{ii}}\,\,
2\pi\delta\left({\omega_1}-{\omega_2}+{\omega_3}-{\omega_4}\right)
\end{equation}

The next terms in the expansion of the interaction
are the triangle terms:
\begin{equation}
\label{trisigns}
{\rm (c),(d)} \,\, = \,\,\mp i\pi\Gamma
\end{equation}
where $\Gamma$ can be either $\Gamma_1$ or $\Gamma_2$
(the suppressed index structure is an
extension of the two-point case,
as we discussed above) and the
arrowless propagator can have an arrow pointing in
either direction.

Diagrams (f) (and a similar one with the arrows reversed) and (g) are the
quartic terms resulting from the expansion of the interaction.
\begin{equation}
{\rm (f)} \,\, = \,\,- i\pi\frac{\Gamma}{4} \quad
{\rm (g)} \,\, = \,\,+ i\pi\frac{\Gamma}{2}
\end{equation}

{\bf Decoration of the Interaction Vertices.}
At the one-loop level, we do not need to consider
terms of higher order than the above. However,
there can still be terms
of arbitrarily high order in $\Gamma$ because there
are quadratic interaction terms. These can be obtained
by grouping all of the quadratic terms together
and inverting them
to obtain the full propagator:
\begin{eqnarray}
\label{fullprop}
\langle {V^\dagger}^{ij,\alpha\beta}_{{\omega_1}{\omega_2}}(q)\;
{V}^{kl,\mu\nu}_{{\omega_3}{\omega_4}}(-q)
\rangle
\,\, &=& \,\,
{D_0}\,
{\delta^{il}}{\delta^{jk}}\,\,\,
{\delta^{\alpha\nu}}
{\delta^{\beta\mu}}\,\,\,
\;2\pi\delta\left({\omega_1}-{\omega_4}\right)\,
2\pi\delta\left({\omega_2}-{\omega_3}\right)
\cr
& & -\,\, 2\pi i{\Gamma_1}
{D_1}{D_2}\,
{\delta^{ij}}{\delta^{il}}{\delta^{jk}}\,\,
{\delta^{\alpha\beta}}
{\delta^{\mu\nu}}\,\,
2\pi\delta\left({\omega_1}-{\omega_2}+{\omega_3}-{\omega_4}\right)
\cr
& & +\,\,2\pi i {\Gamma_2} {D_2}{D_0}\,
{\delta^{ij}}{\delta^{il}}{\delta^{jk}}\,\,
{\delta^{\alpha\nu}}
{\delta^{\beta\mu}}\,\,
2\pi\delta\left({\omega_1}-{\omega_2}+{\omega_3}-{\omega_4}\right)
\end{eqnarray}
Note that the interaction terms are distinguished from the
propagator by the ${\delta^{ij}}$ on the thermal indices
and the overal $\delta$ energy function.
Here, ${z_1} = z - 2{\Gamma_1} + {\Gamma_2} \equiv
z-2{\Gamma_s}$ and ${z_2} = z + {\Gamma_2}\equiv
z-2{\Gamma_t}$, where we have introduced the
singlet and triplet interaction amplitudes,
${\Gamma_s} = {\Gamma_1} - {\Gamma_2}/2$,
${\Gamma_t}= -{\Gamma_2}/2$ by rewriting:
\begin{equation}
{\Gamma_1} {\delta^{\alpha\beta}}{\delta^{\mu\nu}}
- {\Gamma_2} {\delta^{\alpha\nu}}{\delta^{\beta\mu}}
= \left({\Gamma_1}-\frac{1}{2}{\Gamma_2}\right){\delta^{\alpha\beta}}
{\delta^{\mu\nu}} - \frac{1}{2}{\Gamma_2}
{{\bf \sigma}^{\alpha\beta}}\cdot{{\bf \sigma}^{\mu\nu}}
\end{equation}
The inversion is more simply done in the singlet-triplet
decomposition since these matrices are orthogonal. It
can then be re-expressed in the form (\ref{fullprop})
where ${D_0},{D_1},{D_2}$ are given by:
\begin{eqnarray}
{D_0} &=& \frac{1}{-iz\left({\omega_1}{\sigma^{ii}_3} - 
{\omega_2}{\sigma^{jj}_3}\right) + D{q^2}}\cr
{D_1} &=&\frac{1}{-i{z_1}\left({\omega_1}{\sigma^{ii}_3} - 
{\omega_2}{\sigma^{jj}_3}\right) + D{q^2}}\cr
{D_2} &=&\frac{1}{-i{z_2}\left({\omega_1}{\sigma^{ii}_3} - 
{\omega_2}{\sigma^{jj}_3}\right) + D{q^2}}
\end{eqnarray}

Hence, by inverting the sum of all of the quadratic terms
(rather than just the quadratic term coming from
$S_D$) and amputating the propagators entering and leaving
the vertex, we effect the replacement
\begin{equation}
{\rm (b)} \,\, = \,\,2\pi i\left(
-{\Gamma_1}\frac{{D_1}{D_2}}{D_0^2}\,{\delta^{\alpha\beta}}
{\delta^{\mu\nu}} + {\Gamma_2}\frac{D_2}{D_0}\,
{\delta^{\alpha\nu}}{\delta^{\beta\mu}}\right)\,
{\sigma_3^{ii}}\,\,
2\pi\delta\left({\omega_1}-{\omega_2}+{\omega_3}-{\omega_4}\right)
\end{equation}
This can, equivalently, be obtained by
summing the geometric series of figure \ref{decorate}(a).
It is this peculiar property of this theory --
namely that there are quadratic interactions --
which lead to one-loop RG equations which
have terms of all orders in $\Gamma_2$.

\begin{figure}
\centerline{\psfig{figure=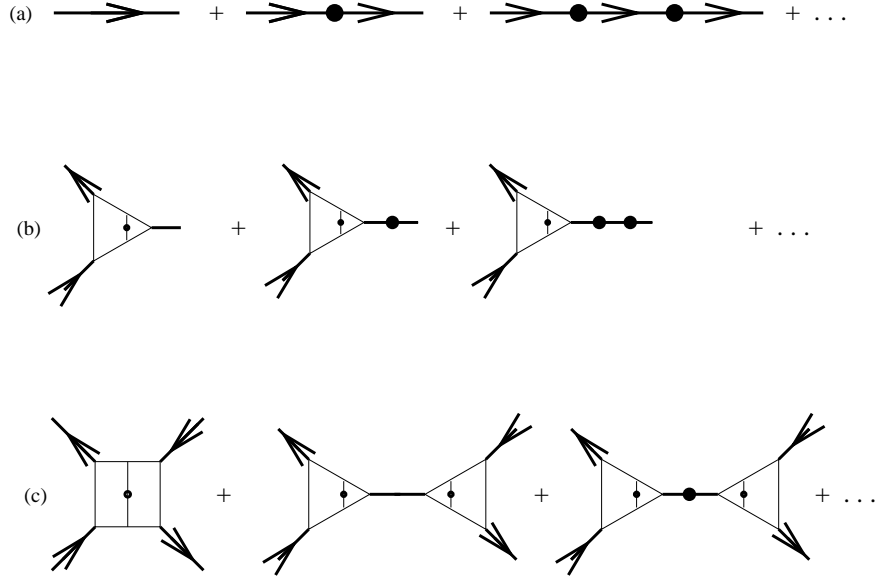,height=3.0in,angle=-90}}
\vskip 0.5cm
\caption{The geometric series of diagrams
which decorate the two-, thre-, and four-point
vertices.}
\label{decorate}
\end{figure}

Similarly, when the interaction in (c), (d), (f),
or (g) is on an internal line,
we can sum the infinite series of figure \ref{decorate}(b) or (c) and,
again, make the replacement
\begin{eqnarray}
\label{replace}
{\Gamma_1} &\rightarrow& {\Gamma_1}
\frac{{D_1}{D_2}}{D_0^2}\cr
{\Gamma_2} &\rightarrow& {\Gamma_2}\frac{D_2}{D_0}
\end{eqnarray}
While the decoration of vertices leads
to a replacement of propagators (\ref{replace}), the
decoration of propagators themselves
-- and, in particular, external legs --
is superfluous since the first term in (\ref{fullprop}),
which has the desired index structure, is not
affected by decoration. The
decoration of an external line of a vertex
amounts to propagator decoration and
is similarly ignored.

\section{Renormalization of the $\sigma$-model}
\label{sec:renorm} 

There are four couplings, $D$, $z$, $\Gamma_{1,2}$,
in our action,
\begin{equation}
S = \frac{1}{2}\int {d^2} x\;
\left[D\;
{\rm tr}\;(\nabla Q^{(1)})^2\;
+\;{\rm tr}\; [(4iz\tilde\Omega-\eta\tilde\Lambda)\; Q^{(2)}]\right]
+
\int d^d x \;
\left[i\pi\Gamma_1 \; Q^{(1)}\gamma_1 Q^{(1)} 
-i\pi\Gamma_2 \; Q^{(1)}\gamma_2 Q^{(1)} \; \right]
\end{equation}
All of these couplings are present in the quadratic terms in the
action. Hence, the RG flows can be computed from the two-point
function alone, separating the different contributions according to
their spin, frequency, and Schwinger-Keldysh  matrix structures.

First, there is the
diffusion constant, $D$ -- or, equivalently,
the resistivity, $g=1/({(2\pi)^2}D)$. In the non-interacting
$\sigma$-model, this is the only coupling constant.
In the interacting case, it is still used as the expansion
parameter: we assume that $g$ is small, but
make no assumptions about $\Gamma_{1,2}$.
The second coupling constant is $z$,
the relative rescaling of time versus space.
$g$ and $z$ are renormalized by the diagrams of figure \ref{gzrenorm}.
The other couplings are the interactions $\Gamma_{1,2}$,
which are renormalized by
the diagrams of figures \ref{expway}-\ref{diag2e}.
Contributions to the $Q$ self-energy
determine the flow of $D$ or $z$
if they are proportional to $q^2$ or $\omega$,
respectively, provided they have the correct thermal
index structure. They determine the flow
of $\Gamma_{1,2}$ if they are independent
of the external and frequency and are diagonal
in the thermal indices; they contribute to one or
the other depending on their spin structure.

It is a matter of taste whether or not we choose
to introduce a `wavefunction'
renormalization\footnote{the quotation marks
refer to the fact that we
mean a renormalization of $V$, not $\psi$}, $\zeta$,
since this can be completely absorbed in
the renormalization of $D$ and $z$. There is
a natural definition of $\zeta$ which is related to
the density of states, but this is moot
for the scaling equations which we consider
because we can rescale our couplings to
eliminate $\zeta$. Hence, we sweep the wavefunction
renormalization under the rug
in this paper.

At tree-level, $D$, $z$, $\Gamma_{1,2}$,
are marginal. (One might have expected
$\Gamma_{1,2}$ to have the dimensions
of $({\rm length})^{-2}$, but they do not
as a result of the extra frequency integral.)
To find the fluctuation corrections
to their scaling behavior,
we use a Wilsonian shell elimination scheme
in which we compute their change
as we integrate out shells
\begin{eqnarray}
\Lambda - d\Lambda < D{q^2} &<& \Lambda, \quad 0<\omega<\Lambda \cr
& &{\rm and}\cr
\Lambda - d\Lambda < \omega &<& \Lambda, \quad 0<D{q^2}<\Lambda
\end{eqnarray}

\begin{figure}
\centerline{\psfig{figure=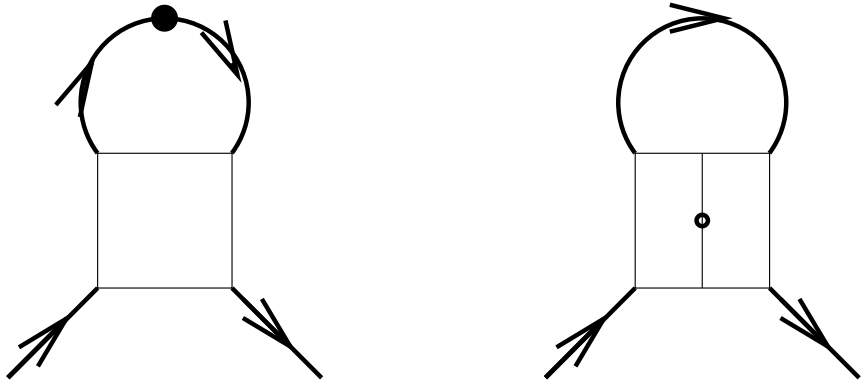,height=2.0in,angle=-90}}
\vskip 0.5cm
\caption{The self-energy diagrams which renormalize
$z$ and $D$.}
\label{gzrenorm}
\end{figure}

{\bf Renormalization of $D$ and $z$.}  Let us examine figure \ref{gzrenorm}.
By inspection, we see that these diagrams have the appropriate matrix
structure for the propagator, i.e.  the matrix structure of (\ref{prop}).
Diagrams without an interaction vertex do not contribute to the
renormalization of $D$ to one-loop order; the reason is that there is a
cancellation between the two (time-ordered and anti-time-ordered)
Schwinger-Keldysh species. In the case of the replica calculation, the
contribution to one-loop order is proportional to the replica number $N\to
0$.

The diagrams in figure \ref{gzrenorm} give the following contribution to the
$Q$-field self-energy, $\Sigma$:
\begin{equation}
\Sigma(q,\Omega+\epsilon,\epsilon) = 
\Sigma_A(q,\Omega+\epsilon,\epsilon)  +
\Sigma_B(q,\Omega+\epsilon,\epsilon) \ ,
\end{equation}
where
\begin{equation}
\Sigma_A(q,\Omega+\epsilon,\epsilon)= 
-4\int\,
\frac{{d^2}k}{(2\pi)^2}\,\frac{d\omega}{2\pi}\,\,
(-2\pi i)\;\left[{\Gamma_1}(k,\omega) - 2 {\Gamma_2}(k,\omega)\right]\,
{D_0}^2(k,\omega)\left[\frac{D}{8}(k+q)^2 - i\frac{z}{8}(\omega+\Omega)\right]\,
\end{equation}
is the contribution of the diagram on the left and
\begin{equation}
\Sigma_B(q,\Omega+\epsilon,\epsilon)= 
4\int\,
\frac{{d^2}k}{(2\pi)^2}\,\frac{d\omega}{2\pi}\,\,
\left(-\frac{\pi i}{4}\right)\;
\left[{\Gamma_1}(k-q,\omega-\Omega) - 2 {\Gamma_2}(k-q,\omega-\Omega)\right]\,
{D_0}(k,\omega)\,
\end{equation}
is the contribution of the diagram on the right.  There is a relative factor
of $2$ between the ${\Gamma_1}$ and ${\Gamma_2}$ contributions because the
former contribution has no free internal spin indices while the latter has
one which is summed over.  Notice that $\Sigma_A$ and $\Sigma_B$ cancel at
$q=\Omega=0$. The contribution of $\Sigma_A$ as well as a piece of
$\Sigma_B$ are absorbed into the wavefunction renormalization.
\footnote{
The $\Sigma_B$ contribution includes a term of the form:
\begin{equation}
4\, {\Gamma_2}\int \frac{{d^2}k}{(2\pi)^2}\,
\frac{d\omega}{2\pi}\,{D_0}(k,\omega) {D_2}(k,\omega)
\end{equation}
which comes from differentiating ${D_0}$ with respect to $q^2$. The
wavefunction renormalization multiplies the $V$ fields and therefore
resurfaces in the renormalization of $\Gamma_2$.} 
Differentiating with respect to $\omega$ or $q^2$, respectively, we get, to
lowest-order in $\Gamma_{1,2}$, the one-loop contributions to $dD$ and $dz$
(the external frequencies and momenta have been
set to zero after differentiation):
\begin{eqnarray}
\label{proprenorm}
dD &=& -{\left(\frac{d\Sigma_B}{dq^2}\right)_{q=0,\Omega=0}}
= -D \; (-2\pi i)\,
\int \frac{{d^2}k}{(2\pi)^2}\,\frac{d\omega}{2\pi}
\,\left[{\Gamma_1}(k,\omega) - 2 {\Gamma_2}(k,\omega)\right]\,
\,{D_0^3}(k,\omega)\,D{k^2}\\
dz &=& -i{\left(\frac{d\Sigma_B}{d\Omega}\right)_{q=0,\Omega=0}}
=  -\frac{1}{2}\int
\frac{{d^2}k}{(2\pi)^2}
\,\left[{\Gamma_1}(k,0) - 2 {\Gamma_2}(k,0)\right]\,
\,{D_0}(k,0)
\end{eqnarray}
In the $dz$ term, the $\Omega$ derivative
cancels the $\omega$ integral.

Making the replacement (\ref{replace}),
we get:
\begin{eqnarray}
\label{Drenormrep}
-\frac{dD}{D} &=&  (-2\pi i)
\int \frac{{d^2}k}{(2\pi)^2}\,\frac{d\omega}{2\pi}\,D{k^2}\,
\left[\,{\Gamma_1}{D_0}(k,\omega){D_1}(k,\omega){D_2}(k,\omega)\,
\,-\,\,2{\Gamma_2}{D_0^2}(k,\omega){D_2}(k,\omega)\right]\\
dz &=& -\frac{1}{2} \left({\Gamma_1} - 2 {\Gamma_2}\right)\,
\int
\frac{{d^2}k}{(2\pi)^2}
\,
\,{D_0}(k,0)
\end{eqnarray}
The contribution to $dz$ is unchanged but the
contribution to $dD$ receives corrections of
all order in $\Gamma_{1,2}$.
This result is the full one-loop result:
it is lowest order in $g$ but
to all orders in $\Gamma_{1,2}$.

The integrals (\ref{proprenorm}) and (\ref{Drenormrep})
are discussed in the appendix. Here, we just state
the result:
\begin{eqnarray}
\frac{dg}{dl} &=& {g^2}\left({\Gamma_1}{f_2}\left(z,{z_1},{z_2}\right)
\,-\,2{\Gamma_2}{f_2}\left(z,z,{z_2}\right)
\right)\cr
\frac{dz}{dl} &=& g\left(-{\Gamma_1} + 2{\Gamma_2}\right)
\end{eqnarray}
where $dl = -d(\ln\Lambda)$ and we have made the substitution
$g=\frac{1}{8\pi D}$ \cite{cond-conv}.
Following Finkelstein, we have introduced the
functions:
\begin{eqnarray}
\label{fdef}
{f_1}\left(z,{z_2}\right) &=& \frac{1}{{z_2}-z}\,
\ln\left(\frac{z_2}{z}\right)\cr
{f_2}\left(z,{z_1},{z_2}\right) &=& \frac{2z_1}{{z_1}-{z_2}}
{f_1}\left(z,{z_1}\right) \,-\,
\frac{2z_2}{{z_1}-{z_2}}
{f_1}\left(z,{z_2}\right)
\end{eqnarray}

As we show in the appendix, the functions $f_{1,2}$
arise as a result of the mismatch between the poles
of the propagators $D_{0,1,2}$. In the non-interacting
case, in which these propagators are equal, these
functions are just constants, ${f_1}\left(z,z\right)=1$,
${f_2}\left(z,z,z\right) = 2$. Interactions split
the energies of  singlet and triplet
particle-hole states from their non-interacting value,
leading to (\ref{fdef}).

{\bf Renormalization of $\Gamma_{1,2}$.}
The diagrams which renormalize $\Gamma_1$
are in figures \ref{expway} and \ref{diag1ab}-\ref{diag1e}
while the related diagrams
which renormalize $\Gamma_2$ are in figures \ref{expway}
and \ref{diag1ab}-\ref{diag1e}. In \ref{diag1ab}-\ref{diag2e},
we have, for the sake of completeness
and pedagogy, enumerated all of the diagrams
which contribute to the renormalization
of  $\Gamma_{1,2}$. On combinatoric grounds, we
see that we have included all $20$ of the diagrams which
contribute to $\Gamma_1$ and the $20$
which contribute to $\Gamma_2$. All of these diagrams must
be taken into account to obtain the correct signs and
relative coefficents of various terms in
the RG equations.
However, these two sets of $20$ are really
elaborations of two sets of $5$ diagrams.
In figure \ref{fivediag}, we give the $5$ basic
diagrams for $\Gamma_1$; figures \ref{diag1ab}-\ref{diag2e} are
put in an appendix to avoid unnecessary
clutter and confusion. From diagram
\ref{fivediag}(a) we obtain the $6$ diagrams
of \ref{diag1ab}(a). From each of \ref{fivediag}(b),
\ref{fivediag}(c), and \ref{fivediag}(d),
we obtain $4$ diagrams shown, respectively,
in figure \ref{diag1ab}(b), \ref{diag1cd}(c),
and \ref{diag1cd}(d). Finally,
figure \ref{fivediag}(e) gives rise to the
two diagrams of figure \ref{diag1e}.

In figure \ref{expway},  $\Gamma_1$ contributes
to the renormalization of  $\Gamma_2$
and vice versa. When only $\Gamma_2$'s
appear in diagrams \ref{diag1ab}-\ref{diag2e}, the spin index structures
dictate that \ref{diag1ab}-\ref{diag1e} contribute
to the flow of $\Gamma_1$ while
\ref{diag2ab}-\ref{diag2e} contribute to the flow of $\Gamma_2$.
If any of the vertices are $\Gamma_1$'s, however,
corresponding diagrams in \ref{diag1ab}-\ref{diag1e}
and \ref{diag2ab}-\ref{diag2e} cancel.
Hence, we need only consider $\Gamma_2$'s
in diagrams \ref{fivediag}-\ref{diag2e}.

\begin{figure}
\centerline{\psfig{figure=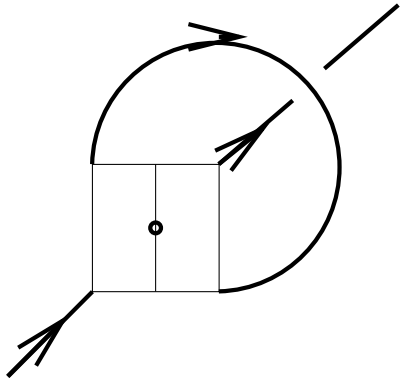,height=2.0in,angle=-90}}
\vskip 0.5cm
\caption{The lowest order diagram which renormalizes
$\Gamma_{1,2}$.}
\label{expway}
\end{figure}

By following the frequencies and spin indices
around the diagram of figure \ref{expway}, we see
that it gives a contribution:
\begin{equation}
d{\Gamma_1}{\delta^{\alpha\beta}}
{\delta^{\mu\nu}}\,-\,
d{\Gamma_2}{\delta^{\alpha\nu}}
{\delta^{\beta\mu}}
 = \left( {\Gamma_1}{\delta^{\alpha\nu}}
{\delta^{\beta\mu}}\,-\,
{\Gamma_2}{\delta^{\alpha\beta}}
{\delta^{\mu\nu}} \right)\,
\int \frac{{d^2}k}{(2\pi)^2}\,
\,{D_0}(k,\omega)
\end{equation}

\begin{figure}
\centerline{\psfig{figure=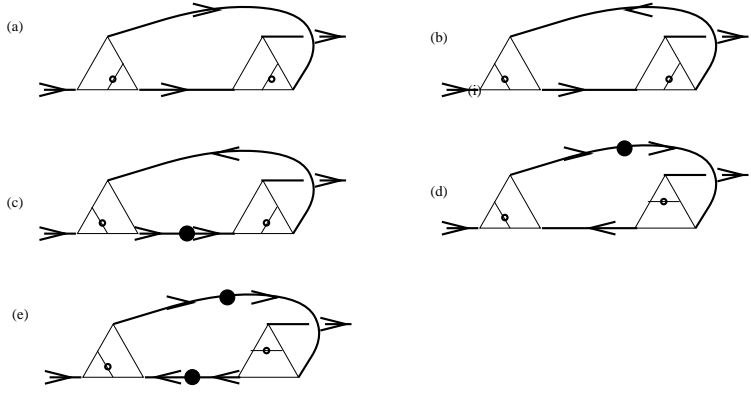,height=2.0in,angle=-90}}
\vskip 0.5cm
\caption{The five `primitive' diagrams which renormalize
$\Gamma_1$. They are elaborated in figures \ref{diag1ab}-\ref{diag1e}. The
corresponding elaboration for $\Gamma_2$ is in figures
\ref{diag2ab}-\ref{diag2e}.}
\label{fivediag}
\end{figure}

Corresponding diagrams of figures \ref{diag1ab}-\ref{diag1e}
and \ref{diag2ab}-\ref{diag2e}
give identical contributions, apart from
a factor of $2$ coming from a spin sum,
so we will focus on \ref{diag1ab}-\ref{diag1e}.
Figure \ref{diag1ab}(a)
gives a contribution
\begin{equation}
-2\pi i\,d{\Gamma_1} =  \; {(i\pi \Gamma_2)^2}
\int \frac{{d^2}k}{(2\pi)^2}\,
\frac{d\omega}{2\pi}\,{D_0^2}(k,\omega)
\end{equation}
The signs of the six diagrams ((iii) and (iv) are two 
diagrams each because the arrow can point
in either direction on the arrowless lines)
are determined by the relative
signs of the two triangle vertices (\ref{trisigns}).
The first two diagrams come with a positive sign
while the other four come with a negative sign.
Adding these together, we obtain the coefficient
$-2$.
Both of the interaction `dots' are on internal
lines, so we make the replacement (\ref{replace}):
\begin{equation}
-2\pi i\,d{\Gamma_1^{(a)}} = -2\; {(i\pi \Gamma_2)^2}
\int \frac{{d^2}k}{(2\pi)^2}\,
\frac{d\omega}{2\pi}\,{D_2^2}(k,\omega)
\end{equation}

In figure \ref{diag1ab}(b), there are four diagrams and they
all come with positive signs.
Only one interaction `dot' is on an internal
line, so we have:
\begin{equation}
-2\pi i\,d{\Gamma_1^{(b)}} = 
4\; {(i\pi \Gamma_2)^2}\int \frac{{d^2}k}{(2\pi)^2}\,
\frac{d\omega}{2\pi}\,{D_0}(k,\omega) {D_2}(k,\omega)
\end{equation}

The diagrams of figure \ref{diag2ab}(b)
vanish because the integrated frequency is
overdetermined. However, the wavefunction renormalization
coming from the neglected part of $\Sigma_B$
makes up the missing contribution.

In figure \ref{diag1cd}(c), the four diagrams have
positive signs. There are three
interactions, but one is on an external line,
so there's a contribution:
\begin{eqnarray}
-2\pi i\,d{\Gamma_1^{(c)}} &=& 
4\; {(i\pi \Gamma_2)^2} \;{(i 2 \pi \Gamma_2)}
\int \frac{{d^2}k}{(2\pi)^2}\,
\int_{0}^{\Lambda}\frac{d\omega_1}{2\pi}
\int_{-\Lambda}^{0}\frac{d\omega_2}{2\pi}
\,{D_0}(k,\omega_1-\omega_2) [{D_2}(k,\omega_1-\omega_2)]^2 \nonumber\\
&=&
4\; {(i\pi \Gamma_2)^2} \;{(i 2 \pi \Gamma_2)}
\int \frac{{d^2}k}{(2\pi)^2}\,
\int_{0}^{\Lambda}\frac{d\omega}{2\pi}
\frac{\omega}{2\pi}
\,{D_0}(k,\omega) [{D_2}(k,\omega)]^2 \ ,
\end{eqnarray}
where we have set $\omega=\omega_1-\omega_2$ as the only variable inside the
propagators, and obtained the phase space factor $\frac{\omega}{2\pi}$ from
integrating the intermediate frequencies.

Figure \ref{diag1cd}(d), on the other hand, comes with a negative sign and
only one internal interaction:
\begin{equation}
-2\pi i\,d{\Gamma_1^{(d)}} =  
-4\; {(i\pi \Gamma_2)^2} \;{(i 2 \pi \Gamma_2)}
\int \frac{{d^2}k}{(2\pi)^2}\,
\int_{0}^{\Lambda}\frac{d\omega}{2\pi}
\frac{\omega}{2\pi}
\,[{D_0}(k,\omega)]^2 {D_2}(k,\omega) 
\end{equation}
The coefficient $4$ is important for the physics of this theory. In
Finkelstein's original paper \cite{Finkelstein}, the contributions
from the diagrams of figure \ref{diag1cd}
were too small by a factor of $2$.  As a result, a metallic fixed
point with infinite conductivity was found. This error was corrected
in \cite{CCLM2,Finkelstein2}.

Finally, we have the two figures \ref{diag1e}(e)
\begin{equation}
-2\pi i\,d{\Gamma_1^{(e)}} =  
-2\; {(i\pi \Gamma_2)^2} \;{(i 2 \pi \Gamma_2)^2}
\int \frac{{d^2}k}{(2\pi)^2}\,
\frac{d\omega}{2\pi}\,
\left(\frac{\omega}{2\pi}\right)^2
\,[{D_0}(k,\omega)]^2 \;[{D_2}(k,\omega)]^2\ ,
\end{equation}
where, as in cases (c) and (d), there are intermediate frequency integrals
that lead to the phase space factor $\left(\frac{\omega}{2\pi}\right)^2$.

The integrals in the expressions above are discussed in the appendix.  Their
upshot is:
\begin{eqnarray}
d{\Gamma_1^{(a)}} &=& -\frac{\Gamma_2^2}{z_2}\,g\,d\left(\ln\Lambda\right)\cr
d{\Gamma_1^{(b)}} &=& 2{\Gamma_2^2}{f_1}\left(z,{z_2}\right)
\,g\,d\left(\ln\Lambda\right)\cr
d{\Gamma_1^{(c)}} &=& -\frac{\Gamma_2^3}{z_2}
{f_2}\left(z,z,{z_2}\right)
\,g\,d\left(\ln\Lambda\right)\cr
d{\Gamma_1^{(d)}} &=& \frac{\Gamma_2^3}{z}
{f_2}\left({z_2},{z_2},z\right)
\,g\,d\left(\ln\Lambda\right)\cr
d{\Gamma_1^{(e)}} &=&  -{\Gamma_2^4}\left(\frac{\frac{1}{z}
+ \frac{1}{z_2} - 2 {f_1}(z,{z_2})}
{\left(z-{z_2}\right)^2}\right)
\,g\,d\left(\ln\Lambda\right)
\end{eqnarray}
Adding these contributions, we obtain:
\begin{equation}
d{\Gamma_1^{(a)}} + d{\Gamma_1^{(b)}} + d{\Gamma_1^{(c)}}
+ d{\Gamma_1^{(d)}} + d{\Gamma_1^{(e)}} \,\,\equiv\,\,-\Phi
= \frac{\Gamma_2^2}{z}\,g\,d\left(\ln\Lambda\right)
\end{equation}

\section{Renormalization Group Equations}
\label{sec:reg}

Gathering these results, we have the flow equations
\begin{eqnarray}
\label{rgeqns}
\frac{dg}{dl} &=& {g^2}\left({\Gamma_1}{f_2}\left(z,{z_1},{z_2}\right)
\,-\,2{\Gamma_2}{f_2}\left(z,z,{z_2}\right)
\right)\cr
\frac{dz}{dl} &=& g\left(-{\Gamma_1} + 2{\Gamma_2}\right)\cr
\frac{d{\Gamma_1}}{dl} &=& g\left({\Gamma_2} +
\frac{\Gamma_2^2}{z}\right)\cr
\frac{d{\Gamma_2}}{dl} &=& g\left({\Gamma_1} +
\frac{2\Gamma_2^2}{z}\right)
\end{eqnarray}

Adding the second and fourth equations, and subtracting
twice the third equation, we find:
\begin{equation}
\frac{d}{dl}\left(z-2{\Gamma_1}+{\Gamma_2}\right) = 0
\label{rgward}
\end{equation}
Although we have only verified (\ref{rgward}) to one-loop,
it is, in fact, an exact relation which follows from
the Ward identity for charge conservation, as we discuss
in appendix C.

Since a constant of integration in (\ref{rgward})
can be absorbed into a rescaling of frequencies,
we can set ${\Gamma_1}=(z+{\Gamma_2})/2$.
Changing variables from $\Gamma_2$ to
${\gamma_2}={\Gamma_2}/z$
we have the equations:
\begin{eqnarray}
\frac{dg}{dl} &=& {g^2}\left(1 +
3\left(1-\frac{1+{\gamma_2}}{\gamma_2}\,
\ln\left(1+{\gamma_2}\right)\right)\right)\cr
\frac{dz}{dl} &=& z\,g\left(-\frac{1}{2}
+\frac{3}{2}{\gamma_2}\right)\cr
\frac{d{\gamma_2}}{dl} &=& g{\left(1+{\gamma_2}\right)^2}
\end{eqnarray}

>From the final equation, we see that 
${\gamma_2}$ increases at long length scales.
This is not problematic since we made no assumptions
about the smallness of $\Gamma_2$. When $\gamma_2$
becomes sufficiently large, $z$ begins to grow,
and $g$, after an initial increase, begins to decrease.
Since these equations are valid in the small $g$
regime, it would appear that the flow improves their
validity. However, $\gamma_2$ and $z$ diverge
at a finite length scale. Since these equations
include all orders in $\gamma_2$, their breakdown
signals the onset of non-perturbative physics.

It is useful, as a comparison, to consider the
BCS interaction in a Fermi liquid.
There, the flow equation for the BCS interaction,
$V$, can be computed to all orders
in perturbation theory:
\begin{equation}
\frac{dV}{dl} = -V^2
\end{equation}
The one-loop RG result is the full story
because there is only a geometric series
of bubble diagrams. This equation also
breaks down at a finite length scale --
the coherence length. At this length scale,
the non-perturbative physics of pairing takes over.
Similarly, the breakdown of our flow equations (\ref{rgeqns})
implies that non-perturbative physics -- such as the
formation of local moments -- determines the
behavior of the system. Such physics cannot
be accessed perturbatively from the diffusive
saddle point.

\section{Discussion}
\label{sec:disc}

In this paper we have shown how the Schwinger-Keldysh dynamical
formulation \cite{CLN1} can be applied in treating a disordered
interacting electronic problem. We reproduce Finkelstein's
Renormalization Group equations for the interaction strength and
conductance. In our approach, the RG procedure is carried out in a
very simple way, in close resemblance to a $\phi^4$ theory (with cubic
terms as well).

For the case of the diffusive Fermi liquid state, the calculation
using Schwinger-Keldysh  is very similar to the replica method. The two thermal
indices for time-ordered and anti-time-ordered propagation play a
similar role to that of the replica indices in the particular case of the
diffusive Fermi liquid regime studied by Finkelstein. The disorder
couples fields with different thermal indices (or replicas), while the
interaction does not mix thermal indices (or replicas). The
Schwinger-Keldysh indices,
like the replicas, act in this example as simple book-keeping
devices. In the replica calculation, the diffusive Fermi
liquid state is replica
symmetric. This is the underlying physical reason for the simplicity
of the replica structure.

There should exist much more interesting saddle points where the
replica symmetry is broken. One such point is the Wigner glass phase,
which occurs in the limit of low electronic densities. As in the cases
with other types of glasses, such as spin glasses, we know that
replica symmetry breaking is connected to broken ergodicity. In the
Schwinger-Keldysh approach we expect that broken ergodicity (and thus replica
symmmetry breaking) will manifest itself in
subtleties related to the $\omega\rightarrow 0$
limit and the symmetry between time-ordered and
anti-time-ordered products. The Schwinger-Keldysh
provides a natural tool to study dynamical effects, which is a
direction that we will explore.

\begin{center}
{\bf ACKNOWLEDGEMENTS} 
\end{center}

We would like to thank Sudip Chakravarty, Eduardo Fradkin,
Patrick Lee, Adrianus M.M. Pruisken, and Michael
Stone for discussions. The authors would like to thank the
Institute for Theoretical Physics, UCSB for hospitality during the
workshops on Low-Dimensional Quantum Field Theory (1997) --
where this work was begun --
and Disorder and Interactions in the Quantum Hall Effect
and Mesoscopic Systems (1998)-- where this work was completed.
This work was supported in part by the NSF under grant
NSF-PHY 94-07194, grant NSF-DMR-94-24511 at the U.
of Illinois (C.C.), and the A.P. Sloan Foundation
(A.W.W.L.).

\appendix

\section{Elaboration of the Five Basic Diagrams}

The five basic diagrams of figure \ref{fivediag} are a
schematic representation of the following
20 diagrams which renormalize $\Gamma_1$.

\begin{figure}
\centerline{\psfig{figure=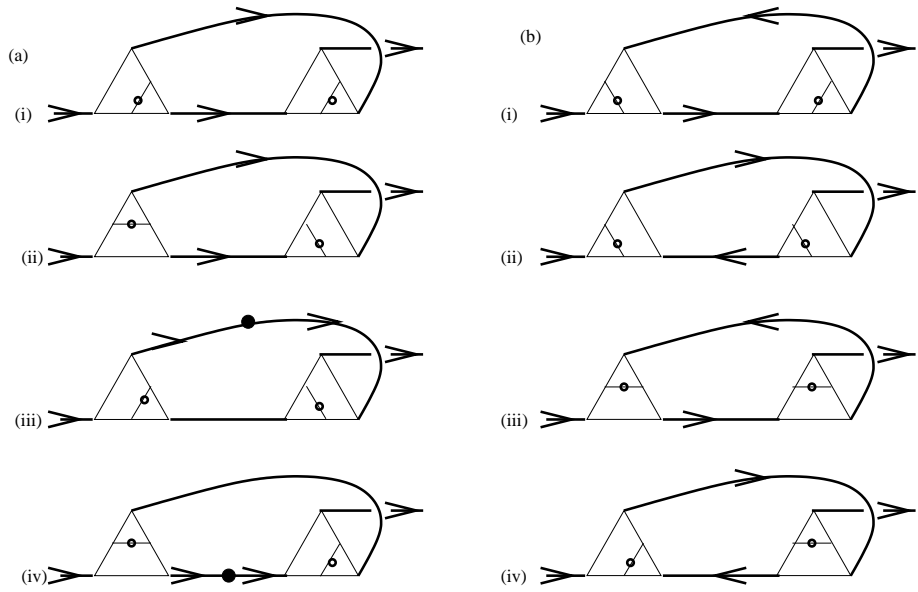,height=3.0in}}
\vskip 0.5cm
\caption{Diagrams (a) and (b) renormalizing $\Gamma_1$.}
\label{diag1ab}
\end{figure}

\begin{figure}
\centerline{\psfig{figure=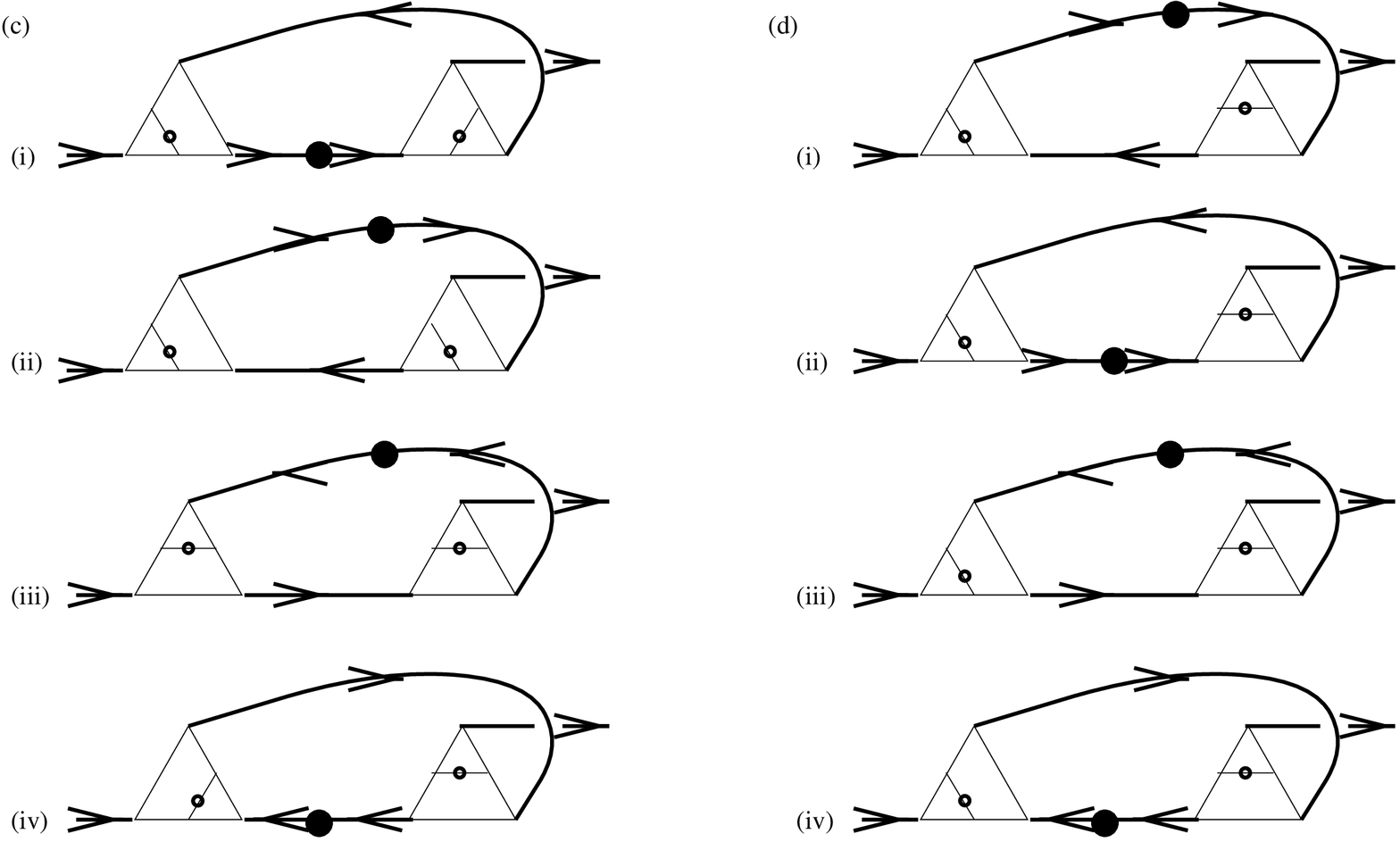,height=3.0in,angle=-90}}
\vskip 0.5cm
\caption{Diagrams (c) and (d) renormalizing $\Gamma_1$.}
\label{diag1cd}
\end{figure}

\begin{figure}
\centerline{\psfig{figure=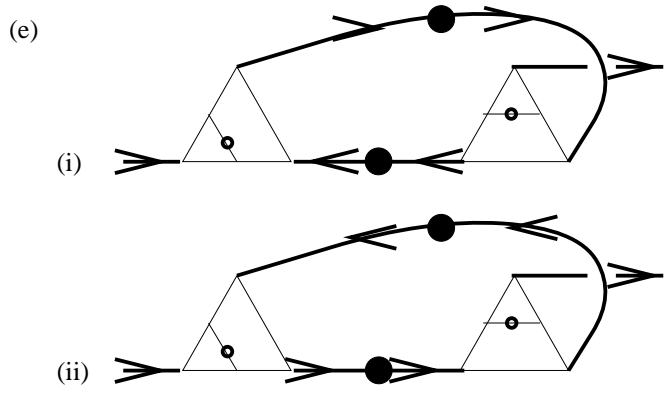,height=2.0in,angle=-90}}
\vskip 0.5cm
\caption{Diagrams (e) renormalizing $\Gamma_1$.}
\label{diag1e}
\end{figure}

The corresponding diagrams which renormalize
$\Gamma_2$ are:

\begin{figure}
\centerline{\psfig{figure=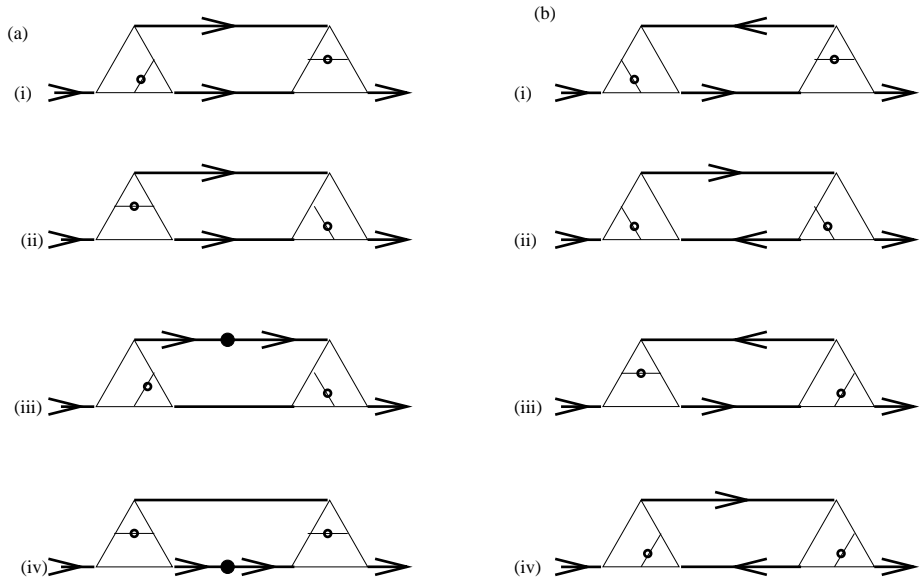,height=3.0in}}
\vskip 0.5cm
\caption{{Diagrams (a) and (b) renormalizing $\Gamma_2$.}}
\label{diag2ab}
\end{figure}

\begin{figure}
\centerline{\psfig{figure=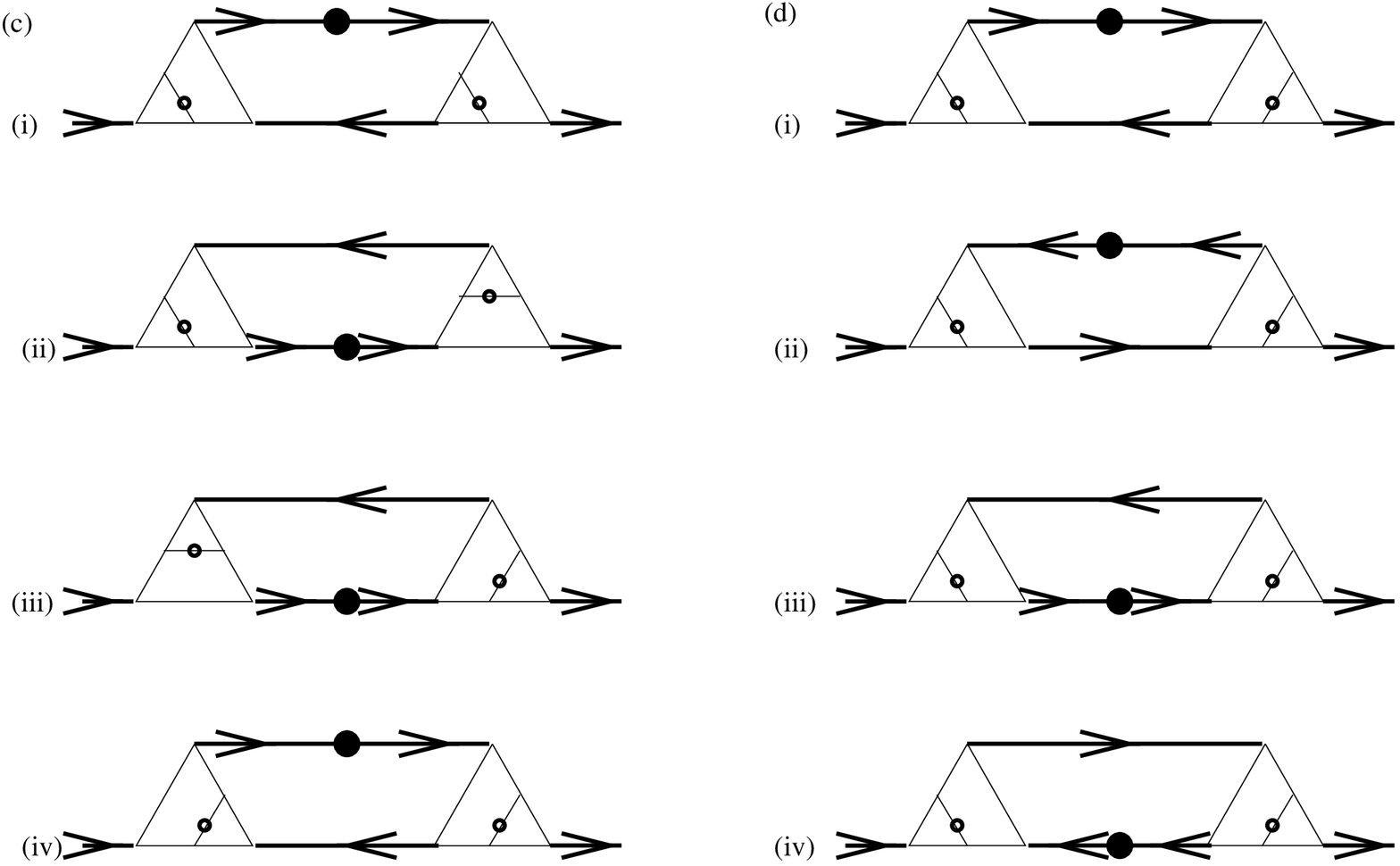,height=3.0in,angle=-90}}
\vskip 0.5cm
\caption{Diagrams (c) and (d) renormalizing $\Gamma_2$.}
\label{diag2cd}
\end{figure}

\begin{figure}
\centerline{\psfig{figure=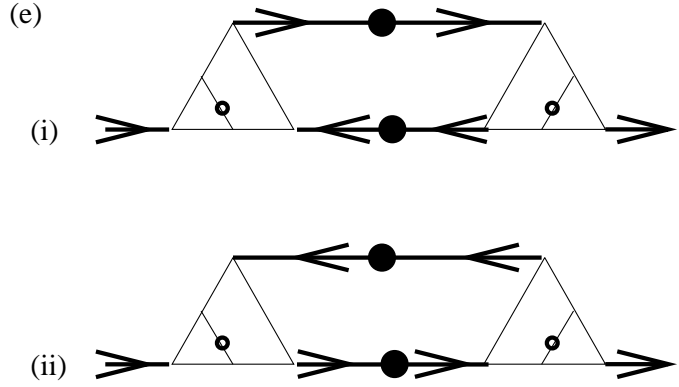,height=2.0in,angle=-90}}
\vskip 0.5cm
\caption{Diagrams (e) renormalizing $\Gamma_2$.}
\label{diag2e}
\end{figure}

\section{Integrals}

The following five logarithmically divergent
integrals arise in the calculation
of the diagram of figure \ref{expway} and diagrams (a)-(e).
We define $x=D{q^2}$ and integrate
over the shells $\Lambda - d\Lambda < x < \Lambda, 0<\omega<\Lambda$
and $\Lambda - d\Lambda < \omega < \Lambda, 0<x<\Lambda$.

In the diagram of figure 4, there are no free frequencies; there is
only a momentum integration.
\begin{equation}
\frac{1}{4\pi D}\,\,\int
\,dx\,\,{D_0}(x,\omega)\,\,=\,\,
\frac{1}{4\pi D}\,\,{\int^\Lambda_{\Lambda - d\Lambda}}
\,\,dx\,\,\frac{1}{x}
\,\,=\,\,\frac{1}{4\pi D}\,\,d\left(\ln\Lambda\right)
\end{equation}

In diagrams (a)-(e), there are multiple propagators with
different $z$'s. By rewriting these products of
propagators as a sum over poles in
$x$ and $\omega$, we reduce them to $d(\ln\Lambda)$ times
non-divergent integrals. These integrals give
non-trivial functions of the $z_i$'s.
\begin{eqnarray}
\frac{1}{2{(2\pi)^2}D}\,\,\int\, dx\,d\omega \,\,
{D_0}(x,\omega)\,{D_2}(x,\omega)\,\,&=&\,\,
\frac{1}{2{(2\pi)^2}D}\,\,{\int^\Lambda_{\Lambda - d\Lambda}} dx\,
{\int_0^\Lambda}d\omega \,\,\frac{1}{x-iz\omega}\,
\frac{1}{x-i{z_2}\omega}\cr
& & \,\,+\,\,
\frac{1}{2{(2\pi)^2}D}\,\,{\int^\Lambda_{\Lambda - d\Lambda}} d\omega\,
{\int_0^\Lambda}dx \,\,\frac{1}{x-iz\omega}\,
\frac{1}{x-i{z_2}\omega}\cr
&=&\frac{1}{2{(2\pi)^2}D}\,\,d\Lambda\,{\int_0^\Lambda}d\omega\,
\frac{i}{\left({z_2}-z\right)\omega}
\left(\frac{1}{\Lambda-iz\omega}-\frac{1}{\Lambda-i{z_2}\omega}\right)
\cr & &\qquad\,\,+\,\,
\frac{1}{2{(2\pi)^2}D}\,\,
\frac{i}{\left({z_2}-z\right)}\,\frac{d\Lambda}{\Lambda}\,
{\int_0^\Lambda}dx
\left(\frac{1}{x-iz\Lambda}-\frac{1}{x-i{z_2}\Lambda}\right)
\cr
&=& \frac{i}{{z_2}-z}\,\ln\left(\frac{z_2}{z}\right)\,\,
\frac{1}{2{(2\pi)^2}D}\,\,d\left(\ln\Lambda\right)\cr
&=&i\, {f_1}\left(z,{z_2}\right)\,\,
\frac{1}{2{(2\pi)^2}D}\,\,d\left(\ln\Lambda\right)
\end{eqnarray}

Similarly,
\begin{eqnarray}
\frac{1}{2{(2\pi)^2}D}\,\,\int\, dx\;d\omega \,\,
x\,{D_0}(x,\omega)\,{D_2}(x,\omega)\,{D_1}(x,\omega)
\,\,&=&\,\,
\frac{1}{2{(2\pi)^2}D}\,\,{\int^\Lambda_{\Lambda - d\Lambda}} dx\,
{\int_0^\Lambda}d\omega \,x\,\frac{1}{x-iz\omega}\,
\frac{1}{x-i{z_2}\omega}\,\frac{1}{x-i{z_1}\omega}\cr
& &\,\,+\,\,
\frac{1}{2{(2\pi)^2}D}\,\,{\int^\Lambda_{\Lambda - d\Lambda}} d\omega\,
{\int_0^\Lambda}dx \,x\,\frac{1}{x-iz\omega}\,
\frac{1}{x-i{z_2}\omega}\frac{1}{x-i{z_1}\omega}\cr
&=& \Biggl(\frac{2i{z_1}}{\left({z_1}-{z_2}\right)\left({z}-{z_1}\right)}
\,\ln\left(\frac{z}{z_1}\right)\cr
& &\,\,-\,\,
\frac{2i{z_2}}{\left({z_1}-{z_2}\right)\left({z}-{z_2}\right)}
\,\ln\left(\frac{z}{z_2}\right)\Biggr)
\frac{1}{2{(2\pi)^2}D}\,\,d\left(\ln\Lambda\right)\cr
&=&\,i\,{f_2}\left(z,{z_1},{z_2}\right)\,\,
\frac{1}{2{(2\pi)^2}D}\,\,d\left(\ln\Lambda\right)
\end{eqnarray}

\begin{eqnarray}
\frac{1}{2{(2\pi)^2}D}\,\int\, dx\;d\omega\,{\int_0^\omega}d\omega' \,\,
\,{D_0}(x,\omega)\,{D_2}(x,\omega)\,{D_2}(x,\omega)
\,\,&=&\,\,
\frac{1}{2{(2\pi)^2}D}\,\,{\int^\Lambda_{\Lambda - d\Lambda}} dx\,
{\int_0^\Lambda}\omega\,d\omega \,\,\frac{1}{x-iz\omega}\,
\frac{1}{x-i{z_2}\omega}\,\frac{1}{x-i{z_2}\omega}\cr
& &\,\,+\,\,
\frac{1}{2{(2\pi)^2}D}\,
{\int^\Lambda_{\Lambda - d\Lambda}} \omega\,d\omega\,
{\int_0^\Lambda}dx \,\,\frac{1}{x-iz\omega}\,
\frac{1}{x-i{z_2}\omega}\frac{1}{x-i{z_2}\omega}\cr
&=& -\frac{1}{2 z_2}\left(\frac{2}{\left(z-{z_2}\right)}
\,\,-\,\,
\frac{2{z_2}}{\left(z-{z_2}\right)^2}
\,\ln\left(\frac{z}{z_2}\right)\right)
\frac{1}{2{(2\pi)^2}D}\,\,d\left(\ln\Lambda\right)\cr
&=&\,-\frac{1}{2 z_2}\,{f_2}\left(z,z,{z_2}\right)\,\,
\frac{1}{2{(2\pi)^2}D}\,\,d\left(\ln\Lambda\right)
\end{eqnarray}

\begin{eqnarray}
\frac{1}{2{(2\pi)^2}D}\,\,\int dx\;d\omega\;\omega^2
& &
{D_0}(x,\omega)\,{D_0}(x,\omega)\,{D_2}(x,\omega)\,{D_2}(x,\omega)
\,\,=\cr & &\,\,
\frac{1}{2{(2\pi)^2}D}\,
{\int^\Lambda_{\Lambda - d\Lambda}} dx\,
{\int_0^\Lambda}{\omega^2}\,d\omega
\,\,\frac{1}{x-iz\omega}\,\frac{1}{x-iz\omega}\,
\frac{1}{x-i{z_2}\omega}\,\frac{1}{x-i{z_2}\omega}\cr
& &\,\,+\,\,
\frac{1}{2{(2\pi)^2}D}\,
{\int^\Lambda_{\Lambda - d\Lambda}} {\omega^2}\,d\omega\,
{\int_0^\Lambda}dx \,\,\frac{1}{x-iz\omega}\,\frac{1}{x-iz\omega}\,
\frac{1}{x-i{z_2}\omega}\frac{1}{x-i{z_2}\omega}\cr
&=& -i\,\left(\frac{\frac{1}{z} + \frac{1}{z_2} - 2 {f_1}(z,{z_2})}
{\left(z-{z_2}\right)^2}\right)\,\,
\frac{1}{2{(2\pi)^2}D}\,\,d\left(\ln\Lambda\right)
\end{eqnarray}

\section{Ward Identity}

In section VI, we found that our one-loop RG
equations imply that:
\begin{equation}
\frac{d}{dl} z = 2\frac{d}{dl}{\Gamma_s}
\label{rgwardagain}
\end{equation}
We would now like to indicate why this
constraint follows from charge conservation.
Let us imagine returning to our disorder-averaged
action ${S_0} + {S_{rand}} + {S_{int}}$ for interacting
electrons.
Let us consider some fixed realization of the disorder.
The Ward identity which follows from charge conservation,
\begin{equation}
{\partial_\mu}\left\langle T\left({j_\mu}\left(y\right)
{\psi^\dagger}\left({x_1}\right)\psi\left({x_2}\right)\right) \right\rangle
= i\delta\left(y-{x_1}\right)
\left\langle T\left({\psi^\dagger}\left({x_1}\right)
\psi\left({x_2}\right)\right) \right\rangle\,-\,
i\delta\left(y-{x_2}\right)
\left\langle T\left({\psi^\dagger}\left({x_1}\right)
\psi\left({x_2}\right)\right) \right\rangle
\label{wardid}
\end{equation}
relates vertex renormalization (the left-hand-side)
to the renormalization of the propagator (the
right hand side). This is more transparent in
momentum space:
\begin{equation}
{\Lambda_\mu}\left(p,0\right) = 
\frac{\partial}{\partial p_\mu}\,\left(
\omega - {\epsilon_p}-\Sigma\left(p\right)\right)
\end{equation}
${\Lambda_\mu}\left(p,0\right)$ is the Fourier
transform of the correlation function with
a current insertion (at zero momentum)
on the left-hand-side of (\ref{wardid}).
$\Sigma$ is the self-energy. We focus on the
$\mu=0$ component of this equation:
\begin{equation}
{\Lambda_0}\left(p,0\right) = 
\frac{\partial}{\partial \omega}\,\left(
\omega - \Sigma\left(p\right)\right)
\label{pwardid}
\end{equation}
The equality between the left- and right-hand-sides
is exemplified by the relation between diagrams
\ref{warddiag}(a) (the dashed line indicates
the current insertion in the correlation function
of the left-hand-side of (\ref{wardid})) and \ref{warddiag}(b).

As we indicate in figure \ref{warddiag}(a), the diagrams
contributing to the renormalization of the
left-hand-side of (\ref{pwardid})
are equivalent to those which determine
the RG flow of $\Gamma_s$.

Meanwhile, the diagrams contributing to the
renormalization of the right-hand-side
of (\ref{pwardid}) renormalize $z$.
One way of seeing this is to note that
the renormalization of the propagator
determines the RG flow of $z$ since the former
is the renormalization of the term
$i\omega{\psi^\dagger}\psi$ in $S_0$
while the latter is the renormalization of the
equivalent term in $S_D$, namely $tr[\omega Q]$.
The factor of $2$
on the right-hand-side results from the
summation over the spin index $\alpha$
in (b). This index is held fixed as an external index in the
$\Gamma_s$ renormalization

In principle, we can derive the same result
for disorder-averaged correlation functions
directly within the $\sigma$-model
by deriving the corresponding Ward identity.
However, this is more cumbersome because the
current operator and the gauge transformation
rules are more complicated in the $Q$-field
language.

\begin{figure}
\centerline{\psfig{figure=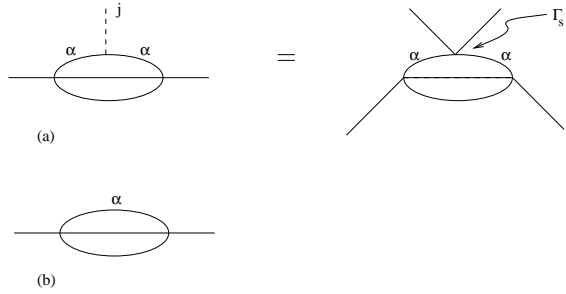,height=1.5in,angle=-90}}
\vskip 0.5cm
\caption{Diagrams contributing to the renormalization
of (a) $\Gamma_s$ and (b) $z$.}
\label{warddiag}
\end{figure}



\begin{thebibliography}{99}
\bibitem{factorization} $Z^{C_3\cup C_4}$ contains
the statistical information such as $F = -T\ln Z^{C_3\cup C_4}$
which is not our main interest here. For this reason
we ignore it.
\bibitem{cond-conv} By rescaling our $Q$ field to set ${Q^2}=1$,
and absorbing the resulting factor of the density of states
$\nu_0$ into $\Gamma_1$, we have essentially set ${\nu_0}=2/\pi$.
Consequently, our conductance is slightly different from
the conventional $g=\frac{1}{{(2\pi)^2}D}$.
\end{thebibliography}
\end{document}